%% file: AAA_Main_01.tex
\documentclass[preprint,12pt,3p,times]{elsarticle}

\usepackage{amssymb}
\usepackage{amsmath}
\usepackage{siunitx}
\usepackage{longtable}
\usepackage{tabularx}
\usepackage{xcolor}
\usepackage{enumitem}
\usepackage{multicol}

\usepackage{url}

\graphicspath{{./figures/}}

\renewcommand{\unit}[2]{{#1}{\,\,#2}}
\setcitestyle{square}

\journal{Journal of Sound and Vibration}

\begin{document}

\begin{tabular}{p{.9\textwidth}}
\hline
$\copyright$ 2023. This manuscript version is made available under the CC-BY-NC-ND 4.0 license\\ \url{https://creativecommons.org/licenses/by-nc-nd/4.0/}\\
The formal publication can be found here:\\
\url{https://doi.org/10.1016/j.jsv.2023.117926}\\
\hline
\end{tabular}

	\begin{frontmatter}
		
		\title{Aeroacoustic testing on a full aircraft model at high Reynolds numbers in the European Transonic Windtunnel}
		
		\author{Thomas Ahlefeldt}
		\author{Daniel Ernst}
		\author{Armin Goudarzi}
		\author{Hans-Georg Raumer}
		\author{Carsten Spehr}
		\address{German Aerospace Center (DLR), 37083 G\"ottingen, Germany}
		
		\begin{abstract}
        This paper presents an end-to-end approach for the assessment of pressurized and cryogenic wind tunnel measurements of an EMBRAER scaled full model close to real-world Reynolds numbers. The choice of microphones, measurement parameters, the design of the array, and the selection of flow parameters are discussed. Different wind tunnel conditions are proposed which allow separating the influence of the Reynolds number from the Mach number, as well as the influence of slotted and closed test sections. The paper provides three-dimensional beamforming results with CLEAN-SC deconvolution, the selection of regions of interest, and the corresponding source spectra. The results suggest that slotted test sections have little influence on the beamforming results compared to closed test sections and that the Reynolds number has a profound, non-linear impact on the aeroacoustic emission that lessens with increasing Reynolds number. Further, sources show a non-linear Mach number dependency at constant Reynolds number but are self-similar in the observed Mach number range. The findings suggest that it is possible to study real-world phenomena on small-scale full models at real-world Reynolds numbers, which enable further investigations in the future such as the directivity of sources.
		\end{abstract}
		
		
	\end{frontmatter}
	
	

    \section*{Nomenclature}
	\input{nomenclature.tex}

	\section{Introduction}
	\input{0_Introduction.tex}
  
	\section{Experimental Setup}\label{sec:setup}
	\input{1_Setup.tex}
	
	\section{Methods}\label{sec:methods}
	\input{2_Methods.tex}

    \section{Results}\label{sec:results}

	\input{3_Results.tex}
    
    \section{Aeroacoustic Analysis}\label{sec:analysis}
	\input{4_Analysis.tex}
	
	\section{Conclusion}\label{sec:conclusion}
	\input{5_Conclusion.tex}

	\section*{Author contributions}
    Conceptualisation and methodology: T.A., D.E., A.G. and C.S; Investigation: T.A., D.E., A.G., C.S. and H.G.R.; Software and formal analysis: T.A. D.E. and A.G.; Data curation: D.E.; Writing---original draft preparation: T.A., D.E., A.G. and C.S.; Writing---review and editing: A.G. and H.G.R.; Visualisation: T.A., D.E. and A.G.; Supervision: C.S.; Project administration: T.A. D.E. and C.S.; Funding acquisition: C.S.
	
	\section*{Acknowledgements}
	The measurements were carried out within the German LuFo V-3 project LoCaRe “Localisation and Characterisation of Flight Relevant Noise Sources on High Lift Systems” (Funding reference number 20A1701A).\\
	Many people from the DLR were involved in this comprehensive project. We would like to thank them all, but cannot mention all their names here. In particular, we would like to thank Stefan Haxter for supporting us in carrying out the measurements, Florian Philipp for improving and maintaining our internal beamforming code SAGAS, and Carsten Fuchs and Tobias Kleindienst for technical support.\\
	We also would like to thank the European Transonic Wind Tunnel for the provided infrastructure and help in performing the measurements. Special thanks go to Ann-Katrin Hensch and Peter Guntermann, who showed great personal involvement in successfully carrying out this project. Thanks also go to EMBRAER for the provision of the model and especially to Alysson Kennerly Colaciti for fruitful discussions.

	\bibliographystyle{elsarticle-num-names}
	\bibliography{AAA_Main_01.bib}

\end{document}

%% file: nomenclature.tex
\begin{longtable}{  l l }
	\textit{Variables} & \\
	$\beta$             & $\beta^2=1-|\mathbf{M}|^2$\\
	$\Delta_{x_{-3dB}}$ & distance to source for a 3 dB decay\\
	$\kappa$            & power scaling weight\\
	$\mu$               & dynamic viscosity\\
	$\rho$              & density\\
	$\sigma$            & standard deviation\\
	$\mathcal{F}, \mathcal{G}, \mathcal{H}$       & generic functions\\
	$a$                 & speed of sound\\
	$b$                 & conventional beamforming source map\\
	$c$                 & constraint function\\
	$\mathbf{C}$        & cross-spectral matrix\\
	$D$                 & characteristic length\\
	$E$                 & Young's modulus\\
	$f$                 & frequency\\
	$\widehat{f}$       & generalized frequency\\
	$\mathbf{g}$        & propagation vector\\
	$\mathbf{h}$        & steering vector\\
	$\mathit{He}$                & Helmholtz number\\
	j                   & imaginary unit\\
	$k$                 & wave number\\
	$m$                 & frequency modification exponent\\
	$M$                 & Mach number\\
	$\mathbf{M}$        & Mach number vector $|\mathbf{M}|=M$\\
	$n$                 & power scaling exponent\\
	$N$                 & number of microphones\\
	$p'$                & acoustic pressure fluctuations\\
 	$p_{stat}$          & static pressure\\
	$q$                 & dynamic pressure\\
	$\mathit{Re}$       & Reynolds number\\
	$\mathit{St}$       & Strouhal number\\
	$s$                 & model scaling\\
	SPL                 & sound power level\\
	$\widehat{\text{SPL}}$ & scaled sound power level\\
	$\mathbf{U}$        & free stream velocity vector $u_\infty = |\mathbf{U}|$\\
	$u_\infty$          & free stream velocity\\
	$w$                 & weighting function \\
	$x,y,z$             & spatial coordinates\\
	$\mathbf{x},\mathbf{y}$ & spatial coordinate vectors\\
	 & \\
	\textit{Subscripts} & \\
	$i,l$ & variable indices\\
	c           & center\\
    max         & maximum\\
    min         & minimum\\
	mic         & microphone\\
	stat        & static\\
	 & \\
	\textit{Abbreviations} & \\
	2D          & two-dimensional\\
	3D          & three-dimensional\\
	A/D         & analogue/digital\\
	CAD         & computer-aided design\\
	CLEAN-SC    & CLEAN based on source coherence\\
	CSM         & cross-spectral matrix\\
	DP          & data point\\
	ETW         & European Transonic Windtunnel\\
	FFT         & fast Fourier transform\\
	NR          & nacelle region\\
	PSF         & point spread function\\
	ROI         & region of interest\\
	SIND        & source identification based on spatial normal distribution\\
	SPL         & sound power level\\
	WLR         & wing leading region\\
	WTR         & wing trailing region

\end{longtable}
\addtocounter{table}{-1}

%% file: 0_Introduction.tex
Wind tunnel testing has a long history in the development, validation, and certification of aircrafts and aircraft components (see e.g. \cite{Herkes1998,Oerlemans2004,Soderman2004}). In recent years, the demand for reliable aeroacoustic testing further increased due to the need for reduced sound emission levels~\cite{Europe2020,Peters2018}. Nevertheless, one main driver of aircraft testing is still optimizing the aerodynamics, and thus reducing fuel consumption, and other economic, or ergonomic aircraft design improvements. Another challenge is the drive concept of the open rotor for passenger aircraft, which is being considered as part of the reduction in fuel consumption, but which leads to more noise~\cite{Farassat2009, Kingan2011}. The resulting designs are driven by the increasing accuracy of computational predictions~\cite{Terracol2005,Smith2022}, which require careful calibration and validation measurements~\cite{stoker2003,Dobrzynski2008, Filippone2014}. Since these tests are performed before the build of actual prototypes, small-scale models must be built to fit in the wind tunnel facilities. One of the various challenges is to observe the model under realistic flight Reynolds numbers since many studies show that acoustic phenomena observed at low Reynolds numbers are vastly different. Thus, pressurized and cryogenic wind tunnel facilities were developed, which drastically increased the range of observable Reynolds numbers~\cite{Quest2000,Paryz2012}.\\

Assuming a symmetric aircraft and symmetric flow, half-models are often used for aeroacoustic measurements~\cite{Ahlefeldt2017, Spehr2019, Beamforming2019}. Compared to full-span models, half-models offer advantages such as a higher achievable Reynolds number and frequency range due to the increased model size and the upper limited measurement frequency~\cite{Underbrink2002}.
On the downside, however, half-models do have an impact on the airflow close to the symmetry axis~\cite{Franz1982}, and the measurement of acoustic directivity perpendicular to the flow direction is limited. Full models are often used for aerodynamic performance tests. Thus, despite the disadvantages of a full model (lower achievable frequency range and Reynolds number), it is advantageous to be able to carry out aeroacoustic measurements parallel (“piggy-back”) to the aerodynamic measurements on the full model. Therefore, the current goal is to study full-span models in slotted and closed test sections. This paper presents the first successful aeroacoustic measurement on a full aircraft model under pressurized and cryogenic conditions at ETW.\\

Under cryogenic conditions, several acoustic and fluid dynamic phenomena occur that have to be understood, corrected, or investigated for the successful evaluation of the experiments. This paper aims to guide the reader through the complete process of designing the experiment and provides the aeroacoustic theory to understand and interpret its results. We present the process of designing a suitable microphone array for the model geometry within the spatial facility constraints in section~\ref{sec:setup}. The large array aperture in comparison to the model scale produces a low depth of field, which in return leads to an increased depth resolution, and permits the use of three-dimensional beamforming focus grids. Further, we discuss the selection of measurement configurations that allow separating the influence of the Reynolds number from the Mach number, as the possibility to alter the fluid temperature and pressure allows their independent variation. By varying temperature and static pressure three different Mach numbers in the range from 0.220 to 0.289 are observed at the constant Reynolds number $10^7$. Further, the influence of closed and slotted test sections is examined, as slotted test sections offer aerodynamic advantages~\cite{Meyer2004}, but their potential aeroacoustic effects on the model or the array are unknown. Section~\ref{sec:methods} presents the necessary tools for the correction and evaluation of the measurements. This includes signal processing, beamforming algorithms with a focus on the geometry, the wind tunnel environment, and also the influence of the flow conditions on the aeroacoustic sources. We discuss Region Of Interest integration methods to obtain spectra and discuss how aeroacoustic source types can be identified by their different self-similarities and scaling laws. Section~\ref{sec:results} presents the experimental results. We present the typical display of 2D source maps and integrated spectra. Then, comparisons of the results to observe the influence of the test section (closed/slotted), the Reynolds number, and the Mach number are presented. Section~\ref{sec:analysis} presents an analysis of the observed phenomena which includes a detailed source type dependent self-similarity analysis. The findings are then summarized in section~\ref{sec:conclusion}.

%% file: 1_Setup.tex
\subsection{Wind Tunnel and Model}\label{sec:setup_geometry}

\begin{figure}
\centering
\includegraphics[width=.5\textwidth]{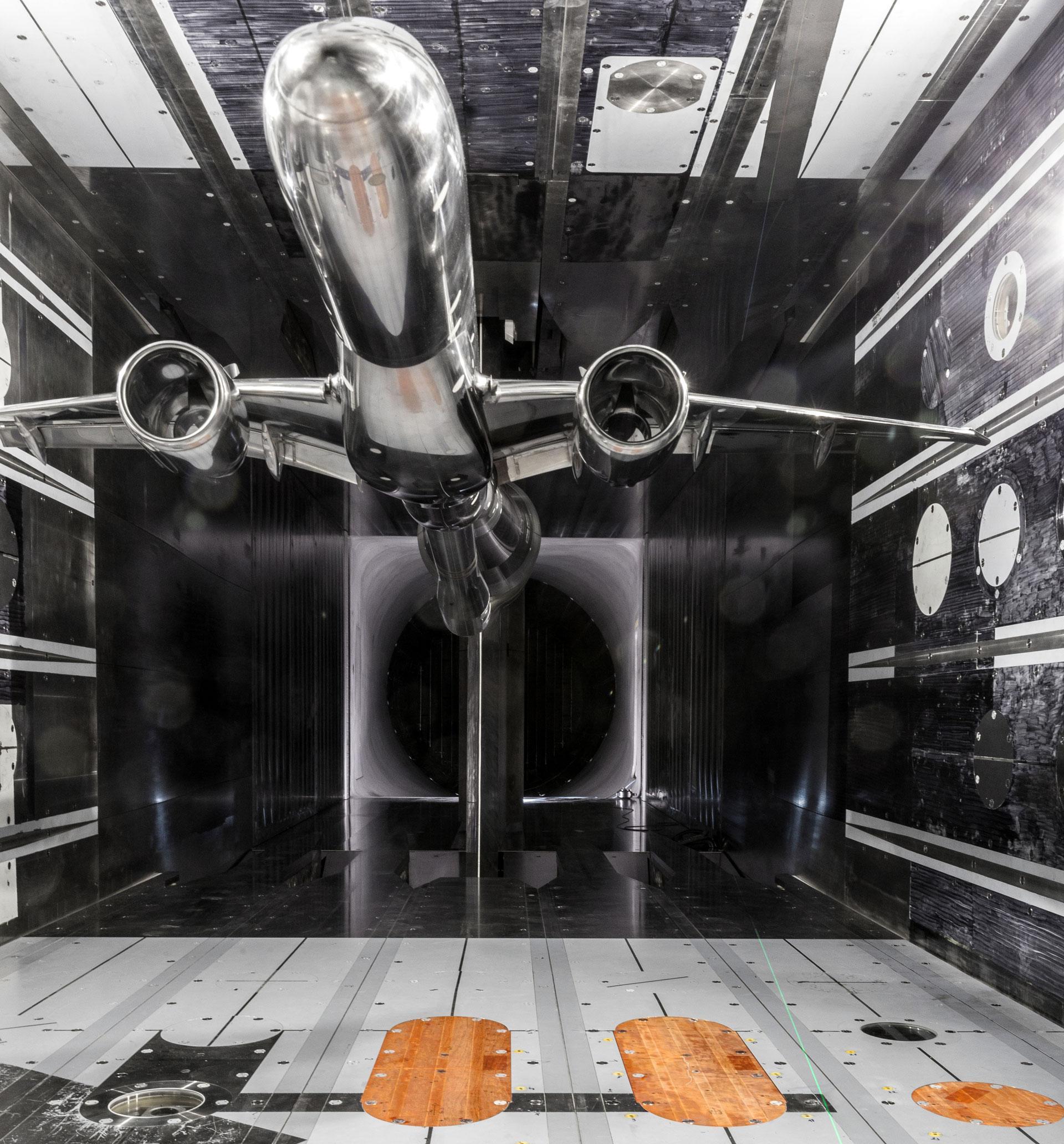}
\caption{Photo of model (top) and microphone array (bottom) in the closed test section. Slot positions are visible as stream wise lines between wooden microphone array inserts. \copyright DLR/ETW}
\label{fig:press_photo}
\end{figure}

The measurements were performed at the European Transonic Windtunnel ETW in Cologne. This facility is a high Reynolds number transonic wind tunnel of G\"{o}ttingen-type. The test section has dimensions of \unit{2.0}{m} x \unit{2.4}{m} and can be operated with closed or slotted walls. The Mach number range is from 0.15 to 1.35. The wind tunnel can be operated at a total pressure range of 115 to \unit{450}{kPa}. Additionally, the temperature can be varied between 110 to \unit{310}{K} by injection of liquid nitrogen, which in total allows an increase in Reynolds number by a factor of approx. 16 compared to ambient pressure and temperature at constant Mach number and model size. Thus, the ETW can provide full-scale Reynolds numbers for scaled models, which is necessary to observe the correct acoustic phenomena and scaling behaviors~\cite{Goudarzi2022}, and offers an independent variation of the Reynolds number from the Mach number, and the load~\cite{Quest2000}.\\

In this study, a full-span aircraft model by EMBRAER is located in the center of the test section as depicted in Figure~\ref{fig:press_photo}. The scaled model is installed in landing configuration, has a mean aerodynamic chord length of about $\SI{0.2}{\metre}$, and is not equipped with landing gear. It is mounted around $\Delta z=\SI{1}{\metre}$ away from the array on a sting that connects its back to a streamlined partition wall in the test section. The pressure side of the model is orientated to the wind tunnels floor where the microphone array is located. During the test, the test section was operated differently, either with slotted floor and ceiling or with closed floor and ceiling.

\subsection{Microphone array and microphone position optimization}\label{sec:setup_optimization}
\begin{figure}
\centering
\includegraphics[width=0.8\textwidth]{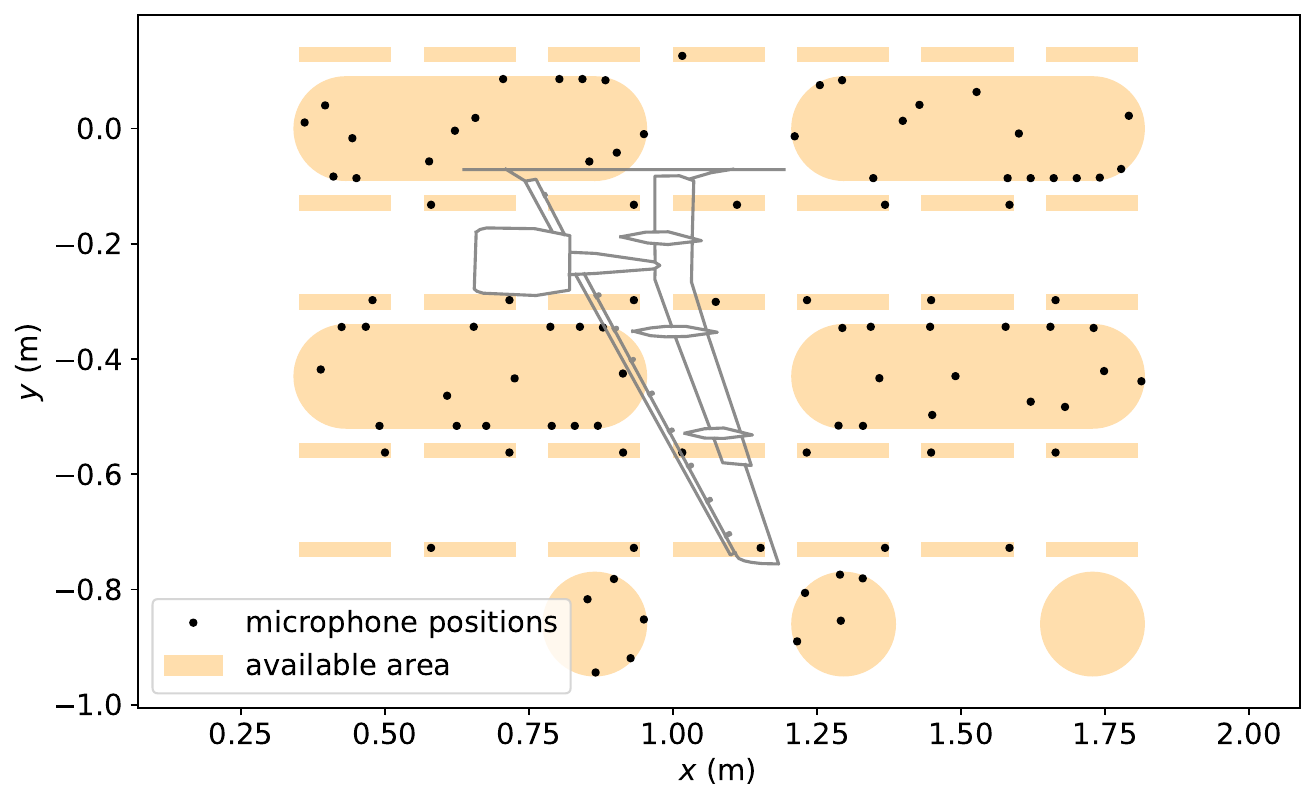}
\caption{Final microphone positions on the floor of the test section. Available areas for the optimization are depicted as colored surfaces and the model's left wing is depicted for reference.}
\label{fig:MicPos_overview}
\end{figure}

Based on the experiences made in a former test~\cite{Ahlefeldt2017}, a microphone array consisting of 96 microphones (Brüel and Kjær cryogenic-type sensor of type 4944A, see section~\ref{sec:miccalib} for calibration details) was designed. Deviating from this previous test, the microphones were placed on the floor instead of on the wall where the available installation area had been significantly extended due to newly manufactured insert slots. The areas available for microphone placement are depicted in yellow in Figure~\ref{fig:MicPos_overview}. For the microphone positioning an optimization problem was set up constrained to the available areas. The objective function to be minimized is the maximum sidelobe of the point spread function (PSF) in the frequency range from 5 kHz to 50 kHz (in 5 kHz steps) for a typical source position one meter above the microphone array.\\ 

More explicitly, the optimization problem can be described as
\begin{equation}
    \min_{\mathbf{x}_\text{mic}}\max_{i,l} w_{i,l}\text{PSF}(\mathbf{x}_\text{mic},f_i,\mathbf{y}_l) \quad \text{ such that } c(\mathbf{x}_\text{mic})\leq 0,
\end{equation}
where $\mathbf{x}_\text{mic} \in \mathbb{R}^{2N}$ are the $x$ and $y$ position of the $N$ microphones, $\text{PSF}(\mathbf{x}_\text{mic},f_i,\mathbf{y}_l)$ is the point spread function, $f_i$ is the $i$-th frequency to be optimized, $\mathbf{y}_l$ the $l$-th focus position and $w_{i,l}$ a weighting factor that depends on frequency and focus position. The chosen frequencies $f_i$ are 5 kHz, 10 kHz, 15 kHz, $\ldots$, 50 kHz, and the focus positions $\mathbf{y}_l$ define an equidistant grid ($\SI{0.5}{\metre} \le x\le\SI{1.4}{\metre}$, $\SI{0}{\metre} \le y\le\SI{0.85}{\metre}$) with a resolution $\Delta x=\Delta y=\SI{2.5}{\milli\metre}$ resulting in 123101 points. These points cover the whole left wing of the model described above.\\

Level set functions are used to describe the spatial constraints of the wind tunnel where microphone placement is possible, as depicted Figure~\ref{fig:MicPos_overview}. These level set functions are summarized in the constraint function $c(\mathbf{x}_\text{mic})$, and in a minimum distance between two microphones of $|\Delta \mathbf{x}_\text{mic}|\ge\SI{30}{\milli\metre}$. The weighting factor $w_{i,l}$ excludes the main lobe from the optimization process and provides lower side lobes in a region around the source position. Due to the continuous optimization problem type, the number of microphones per area must not change and has to be predefined. We used the \textit{fminimax} function from Matlab's Optimization Toolbox~\cite{MATLAB} to solve the optimization problem, which uses gradient-based iterations. Since this optimization problem is far from being convex, an optimal solution can not be guaranteed. The choice of the weighting strategy $w_{i,l}$ and the number of microphones per area were determined by multiple iterations, reviewing the resulting microphone array, and discussing the results with project partners and not by using specific mathematical criteria. In summary, the resulting microphone placement was an iterative process which involved multiple optimizations, discussions, mechanical reviews, and adaption of requirements. The black dots in Figure~\ref{fig:MicPos_overview} show the final, optimized microphone positions.\\

\begin{figure}
\centering
\includegraphics[width=\textwidth]{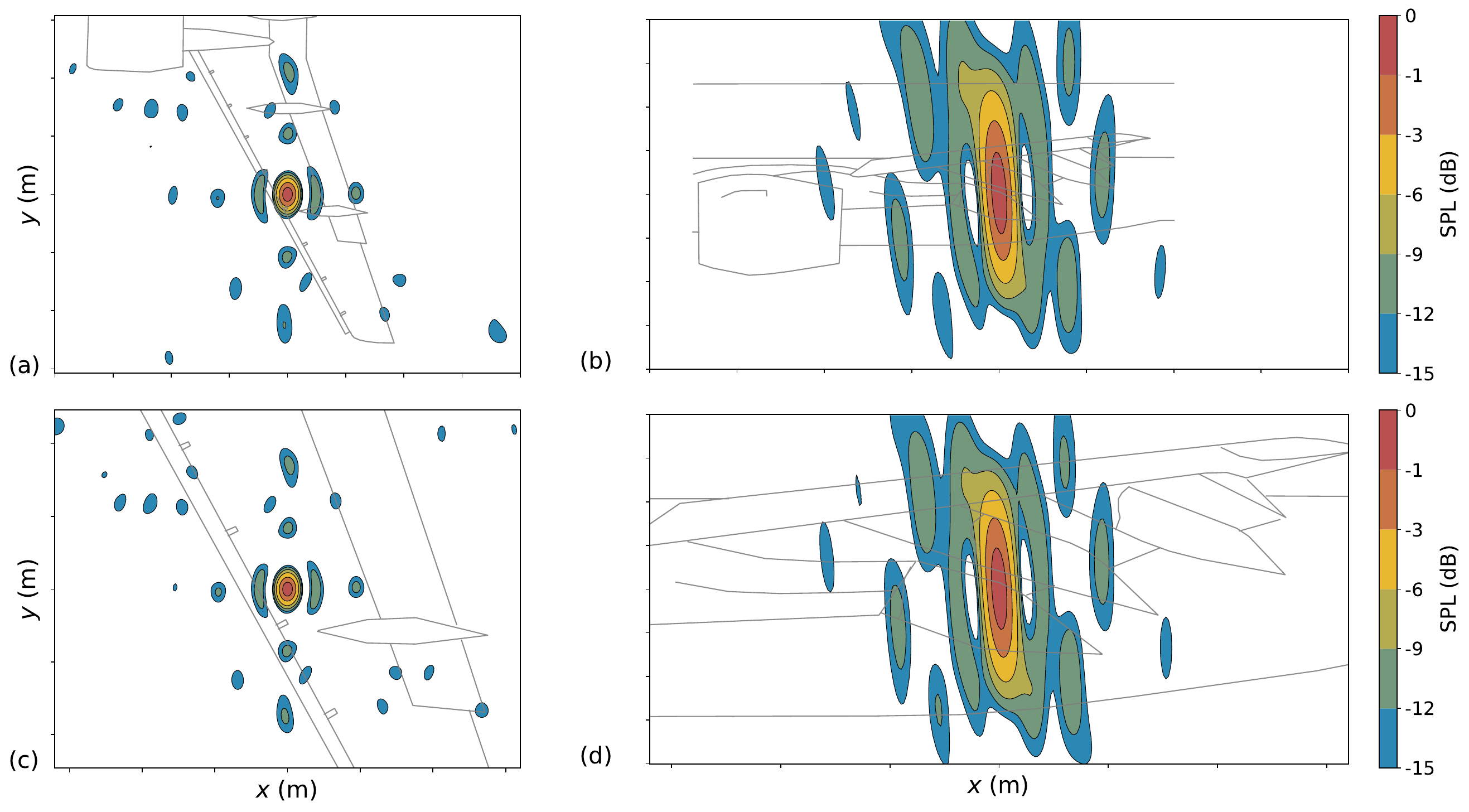}
\caption{2D point spread function for $M$ = 0.22. (a) and (c): $(x,y)$-plane, (b) and (d): $(x,z)$-plane. (a) and (b): $\mathit{St} = 20$, (c) and (d): $\mathit{St} = 50$.}
\label{fig:psf_2d}
\end{figure}

\begin{figure}
\centering
\includegraphics[width=\textwidth]{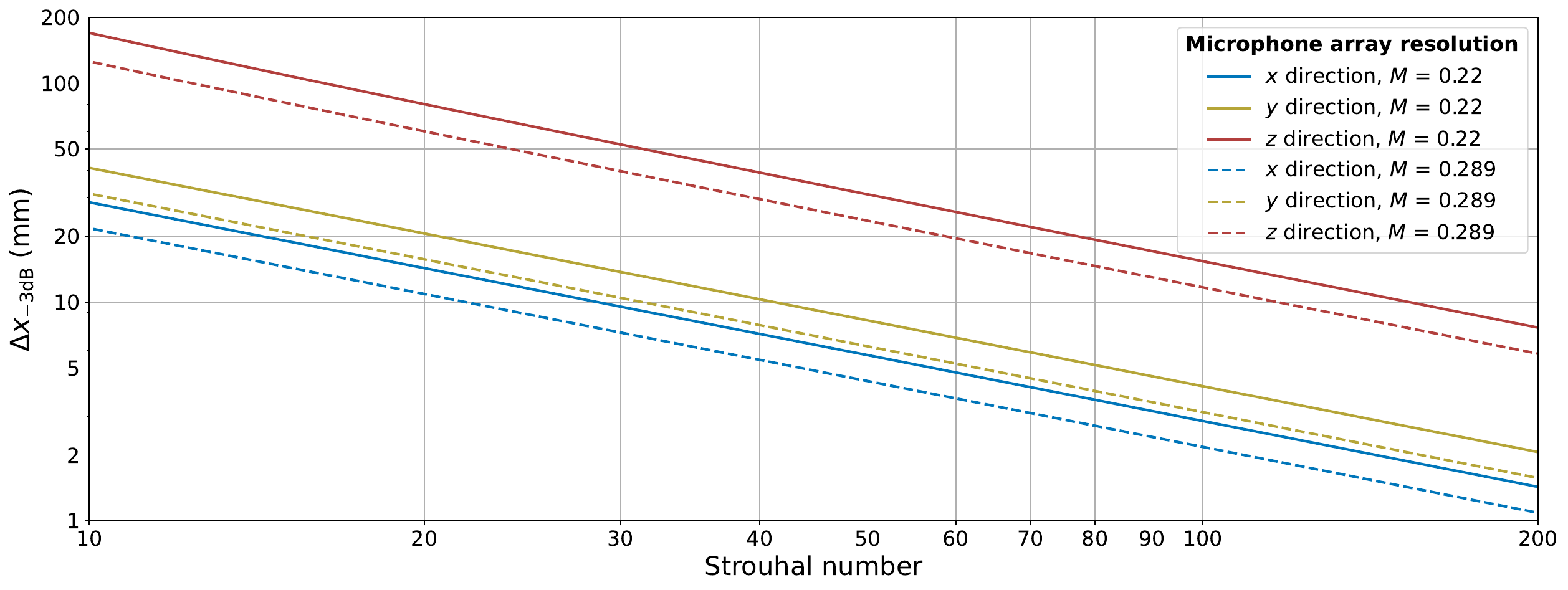}
\caption{Microphone array resolution $\Delta x_{-3 dB}$ for $x$, $y$ and $z$ direction at M = 0.22 (solid lines) and M = 0.289 (dashed lines)}
\label{fig:mainlobe_width}
\end{figure}

Figure~\ref{fig:psf_2d} shows the optimized array's PSF for two different Strouhal numbers (based on the mean aerodynamic chord length, see sec.~\ref{sec:SimLaw}). The main lobe is circular in the $(x,y)$-plane. All side lobes above -15 dB are equidistantly arranged from the main lobe due to the unavailable areas for microphone placing. Especially the side lobes in $y$-direction are much more prominent than in the $x$-direction. In $z$-direction (perpendicular to the microphone array plane) the main lobe is much wider but still achieves a reasonable resolution at $\mathit{St} = 20$ compared to the model size. Figure~\ref{fig:mainlobe_width} shows the array resolution over Strouhal number for different Mach numbers. For Strouhal numbers above 40, a resolution of 10 mm in the $(x,y)$-plane is achieved for all configurations.

\subsection{Data Points}\label{sec:setup_datapoints}
\begin{figure}
\centering
\includegraphics[width=0.7\textwidth]{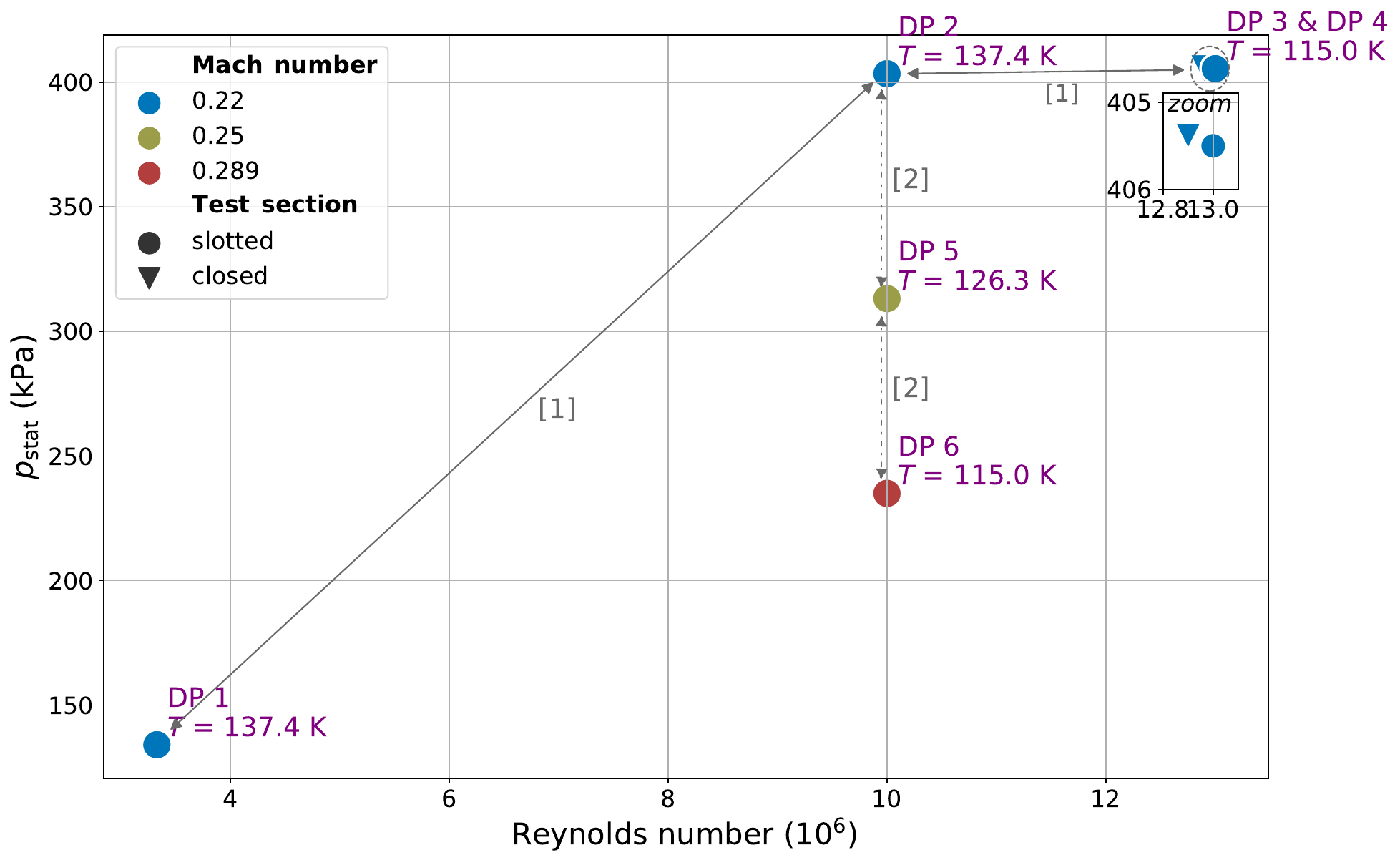}
\caption{Data points versus static pressure and Reynolds number. The double arrows illustrate different comparison possibilities for the analysis.}
\label{fig:datapoints}
\end{figure}

Figure~\ref{fig:datapoints} visualizes the static pressure and the Reynolds number, referenced to the mean aerodynamic chord length, for the measurement configurations. Each data point (DP) is labeled with its temperature, different markers indicate the set up of the test section (slotted or closed floor and top walls). Varying Mach numbers are shown in different colors. The configurations are summarized in Table~\ref{tab:datapoints}.\\

Typically, when altering the temperature and pressure, the corresponding model deformation changes. However, the model's material was chosen in a way that it's temperature dependence (described by Young's modulus $E$) can compensate the increased dynamic pressures $q$ at increased static pressures $p_\text{stat}$. This allows a wide range of constant elastic deformations $q/E$ by combining different temperatures $T$ and static pressures $p_\text{stat}$ to achieve a desired Reynolds number. All displayed data points were measured at the same elastic deformation $q/E=0.0728$.\\

\noindent
\begin{table}
\begin{center} 
\begin{tabular}{ |l||l|l|l|l|l|}
 \hline
 Data &Mach &Temperature &Static pressure &Reynolds &Test \\
 \newline
 point &number &(K) &(kPa) &number ($10^6$) &section \\
 \hline
    DP 1	&0.220	&137.4	&134.10 &03.33 &slotted\\
    DP 2	&0.220	&137.2	&403.45 &10.00 &slotted\\ 
    DP 3	&0.220	&115.0	&405.50 &13.00 &slotted\\ 
    DP 4	&0.220	&115.0	&405.39 &12.90 &closed\\ 
    DP 5	&0.250	&126.3	&313.20 &10.00 &slotted\\ 
    DP 6	&0.289	&115.0	&235.00 &10.00 &slotted\\ 
 \hline
\end{tabular} \end{center}
\caption{Flow parameter of all data points.}
\label{tab:datapoints}
\end{table}

Figure~\ref{fig:datapoints} shows that the data points can be compared in three different ways. They allows studies of Reynolds number dependencies [1] or Mach number dependencies [2] independent of the other parameters (Reynolds number or Mach number, elastic deformation, pressure, temperature). Especially the investigation of the Mach number scaling independent of the Reynolds number should be emphasized, as this is only possible in such a wind tunnel facility. Additionally, the influence of the test section on the results [3] can be investigated.

\subsection{Data acquisition}
Two similar data acquisition systems were used, GBM Viper-48 (16 bit) and GBM Viper-HDR (24-bit). Each system is equipped with its own sigma-delta A/D conversion unit and all A/D converters are synchronized and receive their clocking signal from one common clock source. Both systems possess a second-order high-pass filter with a cutoff frequency of 500~Hz which was used to reduce the influence of the low-frequency wind tunnel noise prior to A/D conversion and to make better use of the dynamic range of the system. Each data point was recorded with a sampling frequency of 250~kHz for a period of \unit{60}{s}.

%% file: 2_Methods.tex
\subsection{FFT Parameter and CSM calculation}
The cross-spectral matrices (CSM) were calculated using Welch's method with a block size of 2500 time samples with 50\% overlap and a Hann window. This results in a frequency resolution of $\Delta f=$~100~Hz and about 12000 spectral averages.

\subsection{Microphone calibration}\label{sec:miccalib}
For the array Br\"{u}el\&Kj{\ae}r microphones of type 4944-W-005 were used. This type of sensor is able to withstand the harsh conditions and, most important, the dependence of its frequency and phase response at the various pressures and temperatures is known~\cite{Ahlefeldt2017}. While their variations in phase response proved to be negligible, the variations in amplitude response for the microphones at different pressures and temperatures were applied to the data.

\subsection{Beamforming}\label{sec:Methods_BF}
For conventional beamforming~\cite[section 2.4]{Sijtsma_NLR_2012} result $b$
\begin{align}
    b = \mathbf{h}^* \mathbf{C} \mathbf{h}
\end{align}
the steering vector $\mathbf{h}$ formulation 4~\cite{Sarradj2012} with diagonal removal 
\begin{align}
\mathbf{h} =\frac{\mathbf{g}}{\left(N(N-1)\left\lVert \mathbf{g} \right\lVert_2^4-\left\lVert \mathbf{g} \right\rVert_4^4\right)^{1/4}}\\
\end{align}
was applied, where $N$ is the number of microphones. The propagation vector $\mathbf{g}$ is derived from the Green's function for the convective Helmholtz equation~\cite[section 3.1]{Raumer2021}. The $l$-th component of $\mathbf{g}$ is given by
\begin{align}
    g_l(\mathbf{y}) = \frac{\exp{\left(-\frac{-\text{j}k}{\beta^2}(-(\mathbf{x}_l-\mathbf{y})\cdot \mathbf{M} + |\mathbf{x}_l-\mathbf{y}|_{\mathbf{M}})\right)}}{4\pi |\mathbf{x}_l-\mathbf{y}|_{\mathbf{M}}}
\end{align}
and depends on the frequency $f$, the speed of sound $a$, the mean flow velocity vector $\mathbf{U}$, the l-th microphone position $\mathbf{x}_l$ and the focus position $\mathbf{y}$:
\begin{align}
    |\mathbf{x}_l-\mathbf{y}|_{\mathbf{M}} &= \sqrt{((\mathbf{x}_l-\mathbf{y})\cdot \mathbf{M})^2 + \beta ^2 |\mathbf{x}_l-\mathbf{y}|^2} \\
    \beta^2 &= 1-|\mathbf{M}|^2\\
    \mathbf{M} &= \frac{\mathbf{U}}{a}\\
    k &= \frac{2 \pi f}{a} \ .
\end{align}
The conventional beamforming maps $b$ are post processed using CLEAN-SC\cite{Sijtsma2007} resulting in the sparse source maps that are presented in section \ref{sec:results}.
For each datapoint, the speed of sound $a$ was derived by averaging over the measurement period by the ETW. Depending on the datapoint, the mean flow angle was calculated by the ETW and included in the mean flow velocity vector $\mathbf{U}$ which was averaged over the measurement period as well. The wind tunnel operates with pure nitrogen and thus avoids any humidity in the fluid. Therefore, the impact of atmospheric damping was estimated to be below 0.4~dB and neglected.

\subsubsection{Geometry}\label{sec:Methods_BF_geometry}
\begin{figure}
\centering
\includegraphics[width=0.7\textwidth]{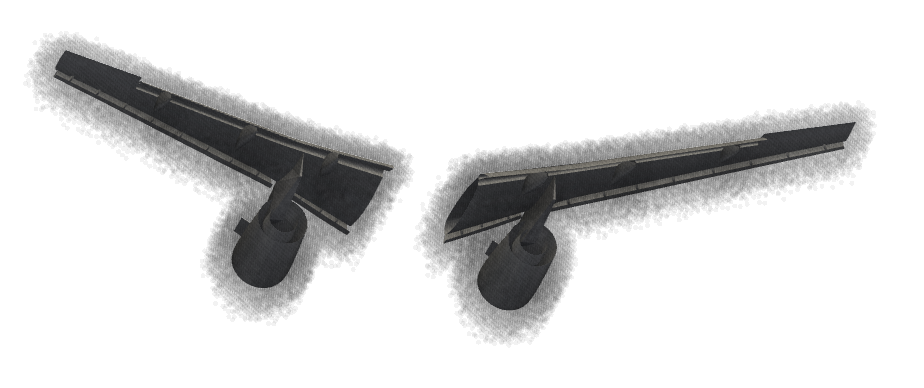}
\caption{Representation of adapted 3D grid covering the hull of the model.}
\label{fig:3DGrid}
\end{figure}

For the beamforming procedure, the geometrical setup is of high importance. The large aperture of the array ($\approx$~1.5~m~x~1.0~m) in comparison to the distance to the model ($\approx$~1.0~m) leads to a very small depth of field (see section~\ref{sec:setup_optimization}). Thus, the slope of the floor ($\approx$~0.55~deg) was also taken into account.\\

The presented model has a high, three-dimensional level of detail. In standard setups, a 2D grid covering the wing with respect to the dihedral angle is often used for the calculations of beamforming results~\cite{Hayes1997,Ahlefeldt2017,Spehr2019,Bahr2021}. Here, the prior discussion of the PSF in section~\ref{sec:setup_optimization} showed that in this test the spatial resolution in $z$-direction has a significant influence on the reconstructed source power. Thus, the source maps need to be evaluated in 3D. In order to avoid unnecessary side lobes with regard to the CLEAN-SC algorithm, the 3D focus grid covers both wings, even if the focus of the analysis lies on the left wing (see section~\ref{sec:setup_optimization}).\\
Figure~\ref{fig:3DGrid} shows the 3D focus grid.\footnote{The 3D CAD model and the 2D sketches used in this paper for the representation of results are roughly simplified generic models and do not correspond to the original model. However, the geometrical key data important for the acoustic analysis are similar.} It covers the wings in an observation volume with equidistant spacing $\Delta x = \Delta y = \Delta z=\SI{1}{\milli\metre}$ leading to a total of 1,288,457~grid points covering the full-span model. The influence of the load is of additional importance which causes the model wings to bend and the potential sound sources to move. Data concerning the airfoil bending were not available to us. Thus, the 3D focus grid was chosen to cover all sources originating from the aircraft model even with bending. This was ensured in particular by a sufficient extension of the focus grid in the $z$-direction.

\subsubsection{Definition of integration regions}\label{sec:methods_subgrids}
\begin{figure}
\centering
\includegraphics[width=0.4\textwidth]{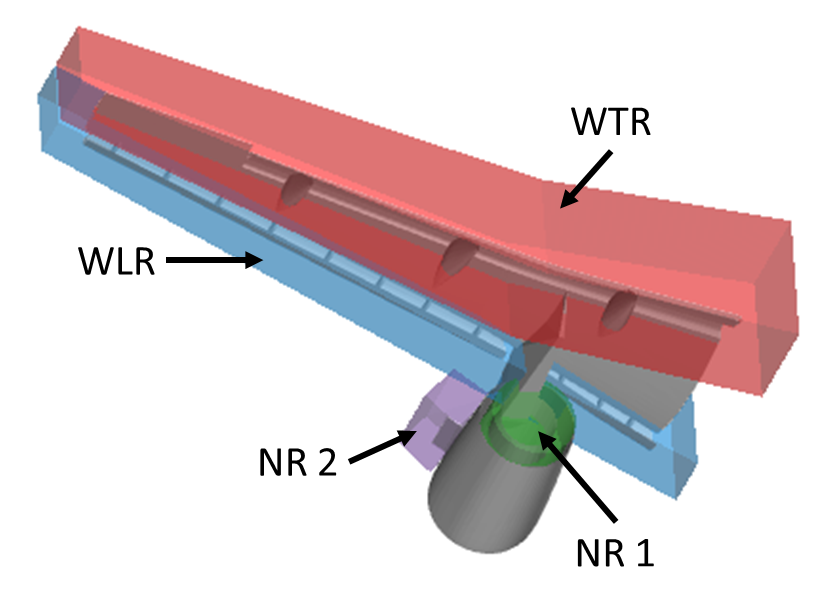}
\caption{Representation of subgrid volumes on the model for the Wing Leading Region (WLR), the Wing Trailing Region (WTR) and two sub-volumes for the Nacelle Region (NR).}
\label{fig:subgrid_3D}
\end{figure}

Beamforming and CLEAN-SC results in four-dimensional maps $\text{SPL}(f,x,y,z)$ that can be spatially integrated to obtain spectra. For the integration it is often beneficial to divide the three-dimensional focus grid into regions of interest (ROI) which can be accounted to typical airframe source regions. Figure~\ref{fig:subgrid_3D} shows the ROI for the Wing Leading Region (WLR), for the Wing Trailing Region (WTR), and two sub-volumes for the Nacelle Region (NR~I and NR~II). In this paper, we will only discuss ROI on the left wing.\\

Often, there exist multiple source types within these typical integration regions~\cite{Goudarzi2022}, which can be spatially distributed or spatially overlapping. To extract individual aeroacoustic source regions from the CLEAN-SC maps without any prior assumptions about their location we use Source Identification based on spatial Normal Distribution~\cite{Goudarzi2021} (SIND), adapted for three dimensions.

\subsubsection{Similarity Laws}\label{sec:SimLaw}
Airframe noise sources can have multiple relevant sound generating mechanisms. Fundamental for the comparison of aeroacoustic sources are the aeroacoustic similarity laws~\cite{Quinlan1996}. According to this theory, aeroacoustic sources are considered similar if the dimensionless quantities Reynolds number ($\mathit{Re}$), Mach number ($M$) and Strouhal number ($\mathit{St}$) or Helmholtz number ($\mathit{He}$) coincide.
\begin{align}
    M &= \frac{u_\infty}{a} \label{eq:M}\\
    \mathit{St} &= \frac{fD}{u_\infty} \label{eq:St} \\
    \mathit{He} &= \frac{fD}{a} \label{eq:He} \\
    \mathit{Re} &= \frac{\rho u_\infty D}{\mu} \label{eq:Re}
\end{align}
In this paper, the mean aerodynamic chord length (about 0.2~m) of the model will be used for the characteristic length $D$. In the following, the scaling of a sound source in terms of the flow characteristics is discussed. A sound source signal can be described as a function of the frequency $f$ and the amplitude $\hat{p}$. 
\begin{equation}
    p' = \mathcal{F}(f,\hat{p}).
\end{equation}
The frequency can be normalized by the speed of sound $a$ and a characteristic length $D$, resulting in a Helmholtz number, or by the Mach number $M$ and $D$, resulting in the Strouhal number. The amplitude can be assumed to be subject to all dimensionless quantities, thus dependent on $u_\infty$, $a$, the density $\rho$ and a characteristic length $D$.\\

Most often, airframe noise sources are modeled as so-called dipole sources~\cite{hubbard1991, Guo2003, Sodermann2004, Dobrzynski2010}. The mean squared acoustic pressure fluctuations of a compact dipole source in the far-field with fixed length scale $D$, Strouhal number and distance $r$ to the source can be written as~~\cite{Ahlefeldt2017}
\begin{equation}\label{eq:p_DP}
    \overline{p'^2}_\mathrm{Dipole} \propto \rho^2 a^4 M^6 \mathcal{G}(\mathit{Re}) \,.
\end{equation}
Here, the dependency on the Reynolds number $\mathcal{G}(Re)$ is included. From the formulation for the acoustic intensity~\cite{Lighthill1952} the acoustic pressure fluctuations of a compact quadrupole source can be derived in a similar way leading to
\begin{equation}\label{eq:p_QP}
    \overline{p'^2}_\mathrm{Quadrupole} \propto \rho^2 a^4 M^8 \mathcal{H}(\mathit{Re}) \,.
\end{equation}
The difference to the dipole formulation is shown by an increased Mach scaling of $M^8$ and a possibly deviating Reynolds number dependency $\mathcal{H}(\mathit{Re})$. Dependent on the comparison to be made, sources are either considered at a fixed Mach number but varying Reynolds number or at a fixed Reynolds number but varying Mach number.

\subsubsection{Reynolds number variation}\label{sec:methods_ReynoldsVariation}
For a fixed Mach number and varying Reynolds numbers equation~\ref{eq:p_DP} and equation~\ref{eq:p_QP} simplify to
\begin{align}
    \overline{p'^2}_\mathrm{Dipole} &\propto \rho^2a^4 \mathcal{G}(\mathit{Re})\,\,\text{and}\\
    \overline{p'^2}_\mathrm{Quadrupole} &\propto \rho^2a^4 \mathcal{H}(Re),
\end{align}
respectively. The decibel correction for dipole and quadrupole sources at different temperatures and static pressures can then be written as (see also~\cite{Ahlefeldt2017})
\begin{equation}\label{eq:decibelcorr}
    \Delta_{\mathrm{SPL}} = 20\mathrm{log}_{10}\left(\frac{\rho a^2}{\rho_0 a_0^2}\right)    
\end{equation}
with $\rho_0=$~1.25~kg/m$^3$ and $a_0=$~337~m/s ($a$ and $\rho$ for pure nitrogen at international standard atmosphere conditions). If this correction is applied to the measured data and assuming the validity of the assumptions made, differences found in the comparison can be attributed to Reynolds number effects (also possibly different for dipole and quadrupole sources). If the assumptions made are violated, differences found can be also related to an effect of deviating source mechanisms (e.g. cavity noise) or noise from non-compact objects. In terms of frequency scaling, the results can be compared versus the Strouhal number, or the Helmholtz number. Due to the constant Mach number both quantities are equivalent.

\subsubsection{Mach number variation}\label{sec:methods_MachVariation}
For varying Mach numbers at constant Reynolds number eq.~\ref{eq:p_DP} and eq.~\ref{eq:p_QP} simplify to
\begin{align}
    \overline{p'^2}_\mathrm{Dipole} &\propto \rho^2a^4 M^6\,\,\text{and}\\
    \overline{p'^2}_\mathrm{Quadrupole} &\propto \rho^2a^4 M^8.
\end{align}
The decibel correction (equation~\ref{eq:decibelcorr}) can be applied to the data but the respective power scaling with the Mach number remains. Thus, given the assumptions of compact sources, differences found in the comparison can be attributed to Mach number effects independent on the Reynolds number.\\

As for the Reynolds number variation source spectra can be compared versus the Strouhal and Helmholtz number. However, due to the increasing Mach number they are not equivalent. Since the Strouhal number often depends on the convective Mach number $M_C$, which is unknown and can differ from the known free-flow Mach number, the observed sources often neither perfectly scale over the free-flow Strouhal and Helmholtz number. This phenomenon was already observed for wind tunnel measurements~\cite{Goudarzi2022}. The authors propose a generalized frequency $\widehat{f}$
\begin{equation}\label{eq:generalized_frequency}
    \widehat{f}=\frac{fD}{M^m a}
\end{equation}
with the modification exponent $m$, that can express a Helmholtz number for $m=0$, a Strouhal number for $m=1$, and everything in-between. A modification exponent $m\le0.5$ indicates a Helmholtz-like scaling, and $m>0.5$ indicates a Strouhal-like scaling behavior of the sources. Note, that the correct modification exponent can only be obtained, if the source spectra contain a single source mechanism, which requires a source type dependent ROI integration, see Section~\ref{sec:methods_subgrids}. The optimal modification exponent is computed by maximizing the self-similarity between spectra at different Mach numbers (cf.~\cite{Goudarzi2022} for a detailed description).\\

The power scaling is then defined by the Mach number's exponent $n$, and results in a scaled level $\widehat{\text{SPL}}$ over the generalized frequency $\widehat{f}$
\begin{equation}\label{eq:Mach_power_scaling}
    \widehat{\text{SPL}}(\widehat{f}) = \left\langle \text{SPL}(\widehat{f},M_i) - 10\log_{10}\left(\frac{M_i}{M_0} \right)^n \right\rangle_i \,.
\end{equation}
Here the angle brackets indicated that we take the average with respect to the different Mach numbers, indexed by $i$. The $\widehat{\text{SPL}}$ is scaled with the given power exponent $n$ from the given Mach number to the reference Mach number $M_0$. For a fixed modification exponent $m$, the power exponent can be estimated by minimizing the weighted sample standard deviation $\sigma$ between the spectra~\cite{Goudarzi2022}
\begin{equation}\label{eq:Mach_scaling_exponent}
\min\limits_{n} \sum_{l}\bigg[\sigma_{i} \big(\text{SPL}(\widehat{f}_l,M_i)-n10\log_{10}(M_i)\big) \big\langle \text{SPL}(\widehat{f}_l,M_i)\big\rangle_{i}^\kappa\bigg]\,.
\end{equation}
Here $\sigma_i$ indicates that we take the sample standard deviation with respect to the different Mach numbers, indexed by $i$. The weighting exponent $\kappa \geq 1$ results in a preference of high SPL in the minimization. The power and frequency scaling can then be used to differentiate between different source mechanisms, and to indicate dipole and quadrupole sources.\\

%% file: 3_Results.tex
\subsection{Reconstructed sources}
This section provides an overview of the beamforming results using the example of DP~2 in terms of a 3D source map, 2D source maps and an integrated spectrum. The labeling of the sources is based on their spatial occurrences but not on their physical source mechanisms or sound generating object. The naming does not imply a causality between the label and the source mechanism.

\subsubsection{Three-dimensional source distribution}
\begin{figure}
\centering
\includegraphics[width=0.7\textwidth]{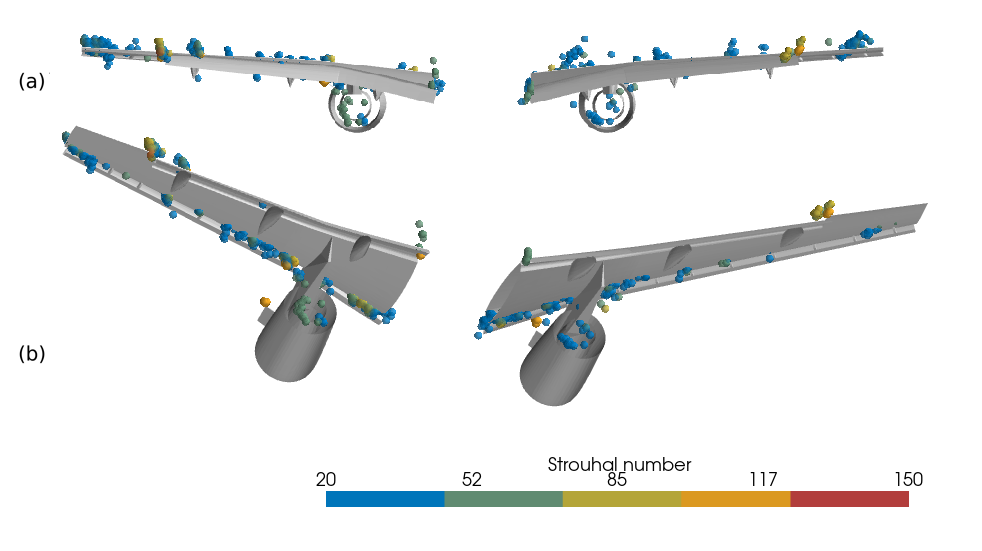}
\caption{Overview of dominant sources on the model for Strouhal numbers from 20 to 150 for DP~2. (a): rear view, (b): bottom rear view. The color represents the Strouhal number.}
\label{fig:3DMap_overview}
\end{figure}

The presented source map was calculated using the beamforming and CLEAN-SC method (see section~\ref{sec:Methods_BF}). Figure~\ref{fig:3DMap_overview} gives an overview of the reconstructed sources on the three-dimensional representation of the wind tunnel model from two different viewing angles. In order to facilitate the overview of the source distributions around the model, only dominant sources with 5~dB dynamic per frequency band are shown. Thus, the sources are color-coded with the Strouhal number (based on the mean aerodynamic chord length, see section~\ref{sec:SimLaw}) instead of the usual source level color-coding. For this data point the Strouhal number range of 20 to 150 corresponds to a frequency range that is relevant for full scale.\\

Sources on the model are observable over the whole Strouhal number range. For the lower Strouhal number range most of the sources are distributed over the Wing Leading Region (WLR), especially in the vicinity of the inboard region, the region close to the wing tip. The mid Strouhal number range covers sources in the Wing Trailing Region (WTR), the Nacelle Region (NR), the fuselage close to the inboard WTR and the outboard WTR. In the high Strouhal number region sources in the the outboard WTR and the NR are dominating.\\

All sources appear to be slightly above the wing with increasing distance to the wing root, since we present the results on the non-deformed wing. Nevertheless, the extended three-dimensional focus grid also captures sources in the near vicinity of the model. The arrangement of the sources also appears slightly different between the left and the right wing. On the right wing the sources, especially the upper sources on the suction side of the wing, seem to be slightly more tilted than on the left side. This appearance is expected by the array positioned under the left wing, see section~\ref{sec:setup_geometry}. In the following we will only present results from the left wing.\\

\subsubsection{2D integrated source maps}
\begin{figure}
\centering
\includegraphics[width=\textwidth]{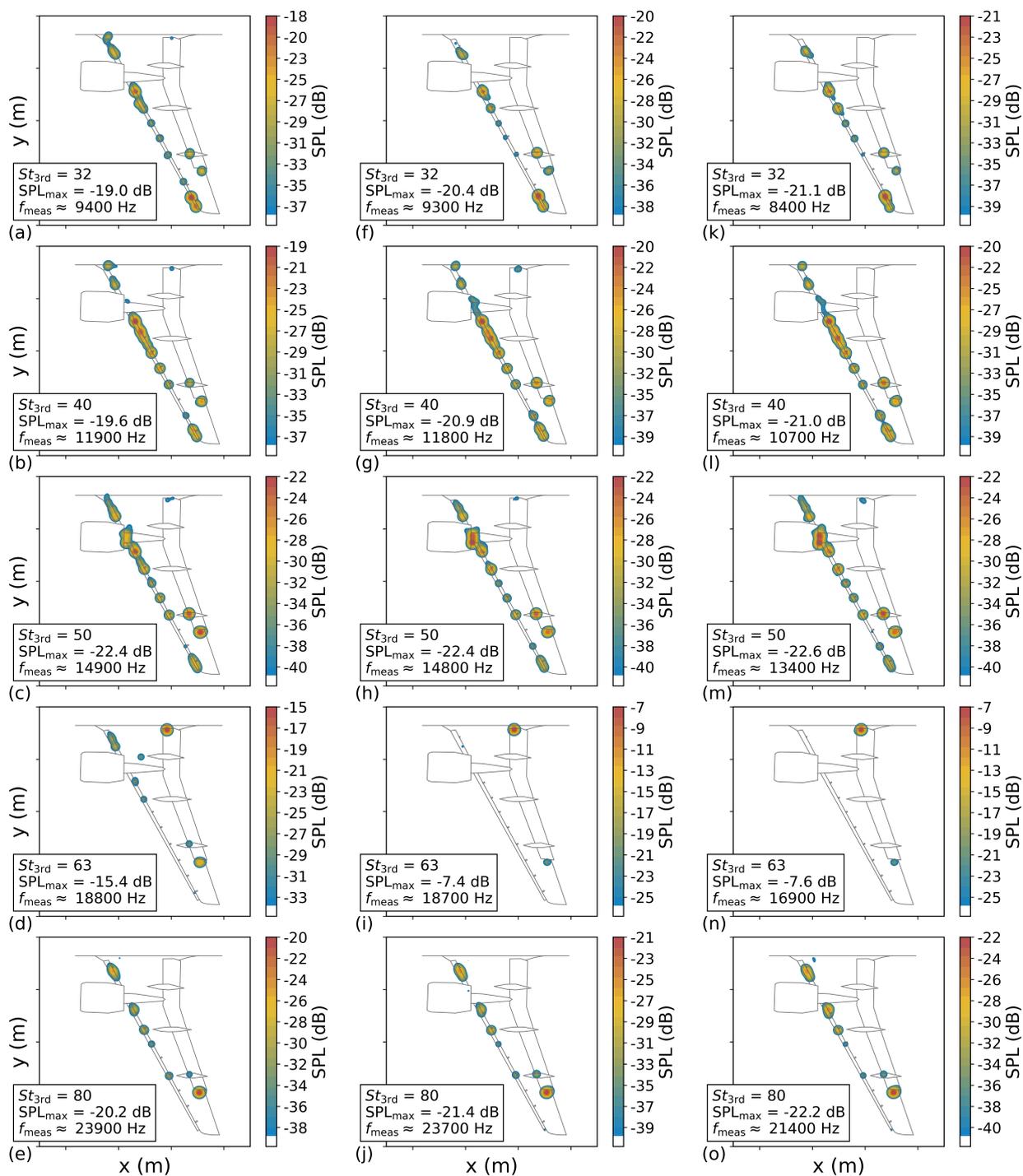}
\caption{Comparison of over $z$ and 3rd octaves integrated source maps for DP~1: (a) to (e), DP~2: (f) to (j), and DP~3: (k) to (o), for $\mathit{St}_\text{c}$ numbers (a), (f), (k): 32, (b), (g), (l): 40, (c), (h), (m): 50, (d), (i), (n): 63, (e), (j), (o): 80.}
\label{fig:Maps_vgl_DP1vsDP2vsDP3}
\end{figure}
In comparison to a 3D representation, a two-dimensional representation of the data in a (cleaned) beamforming map allows for a clear unambiguous presentation. However, an evaluation of the data using a two-dimensional focus grid would have led to a significant underestimation of out-of-focus source powers due to the aforementioned narrow depth of field. Thus, for better comparison of different source maps the reconstructed source powers obtained by CLEAN-SC are integrated along the $z$-direction. All results are corrected regarding the amplitude response of the microphones (section~\ref{sec:miccalib}) and the acoustic conditions (equation~\ref{eq:decibelcorr}). The reference of all source levels was set to an arbitrary reference, here based on the highest SPL of all integrated spectra.\\

Figure~\ref{fig:Maps_vgl_DP1vsDP2vsDP3} shows the integrated CLEAN-SC source maps at the third-octave Strouhal number bands of $\mathit{St}_\text{c} = 32, 40, 50, 63, 80$ with a dynamic range of 20~dB for DP~1, DP~2, and DP~3. The third-octave Strouhal number bands are calculated by integration of the sound power within the third-octave band boundaries for a center Strouhal number $\mathit{St}_\text{c}$:
\begin{equation}
    \mathit{St}_\text{c} = 2^{1/6} \; \mathit{St}_\text{min} = \frac{\mathit{St}_\text{max}}{2^{1/6}}, 
\end{equation}
where $\mathit{St}_\text{min}$ is the lower boundary and $\mathit{St}_\text{max}$ the upper one.\\

As for the 3D map, at low Strouhal numbers sources in the WLR are dominating whereas at higher Strouhal numbers the outboard WTR becomes more dominant. Also the sources in the NR at a Strouhal number of 50 can be identified. Caused by the increased dynamic range compared to the 3D map, additional sources can be found. Next to minor regularly distributed WLR sources, another distinct source in the outer WTR can be found spread over a Strouhal number range of 20 to 100.


\subsubsection{Integrated source spectrum}\label{sec:results_sourcespectrum}
\begin{figure}
\centering
\includegraphics[width=0.7\textwidth]{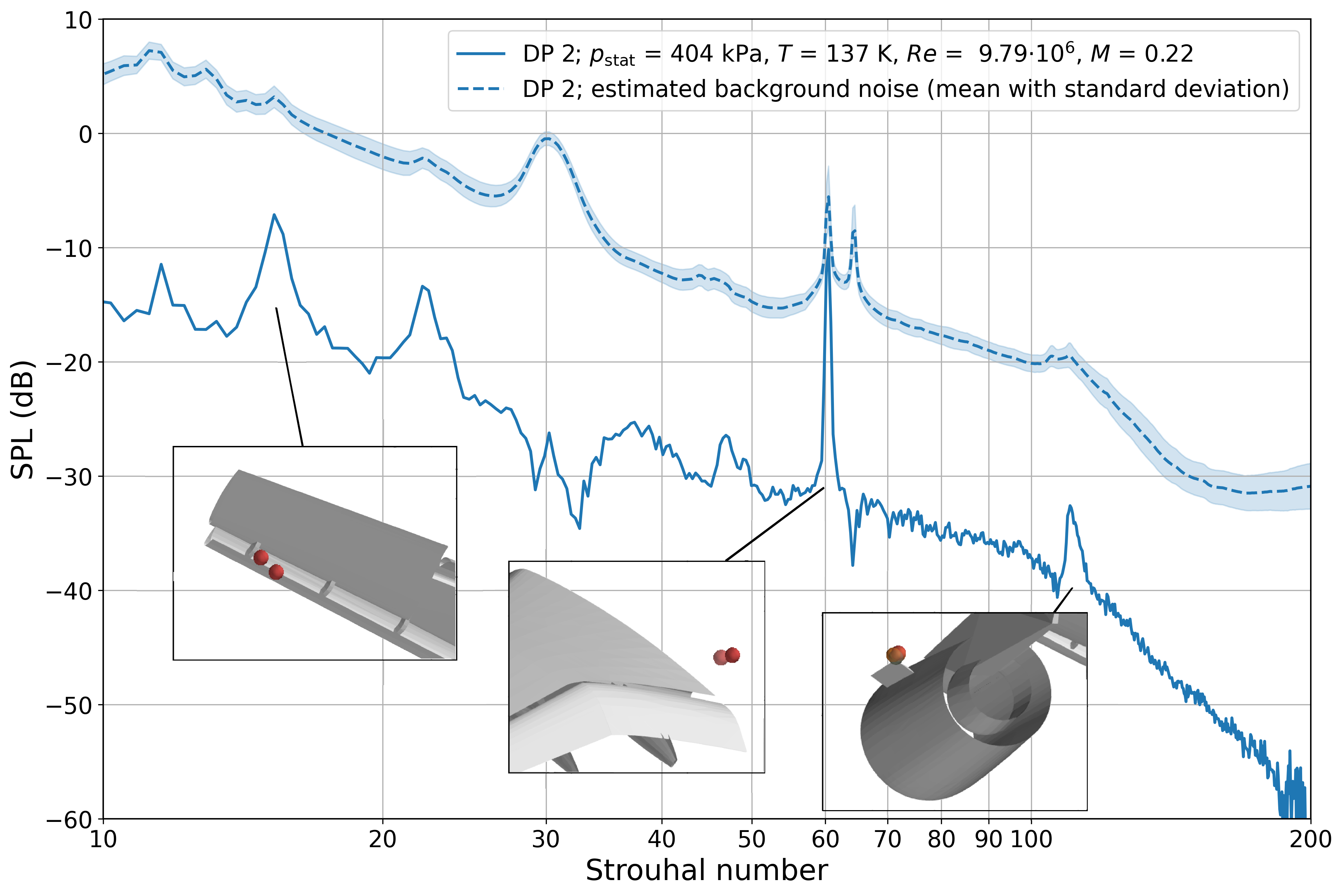}
\caption{Integrated sources of the left wing and background noise vs Strouhal number for DP~2.}
\label{fig:Spec_DP2}
\end{figure}

Figure~\ref{fig:Spec_DP2} provides a closer look at the spectra integrated over the CLEAN-SC beamforming results and the associated source position for several peaks. Additionally, the estimated background noise is shown by averaging the main diagonal of the cross-spectral matrix. Thereby, the background noise contains boundary-layer-induced noise as well as acoustic noise radiated from the model and the wind tunnel. Because the acoustic noise radiated from the model is mostly a minor part of the main diagonal, it is used to represent the background noise. However, this is only an estimate, useful for the following discussion, and does not reflect the unknown true level which should be measured with a clean configuration or an empty test section for each wind tunnel condition.\\

The spectrum shows a typical negative slope decreasing the source level approx. 50~dB between the Strouhal number $0\ge St \ge 200$ (approx. 10~dB per octave). It has a multitude of tonal and conspicuous components. The most striking structures are the following. First, peaks at $\mathit{St}=$~11, 15 and 22 can be attributed to single source positions in the outer WLR. Second, by analyzing the measured data together with the ETW, the drop-in sound pressure level at $\mathit{St}=$~30 could be attributed to an interfering noise source from the wind tunnel test section. Third, a very narrow peak at $\mathit{St}=$~61 can be attributed to a model cavity on the fuselage close to the wing root. Fourth, the peak at $\mathit{St}=$~110 can be attributed to the NR. Note, that the narrow peak at $\mathit{St}=$~61 is not present at DP~4, because the unwanted cavity was fixed during the measurement campaign. Also visible are two less dominant peaks around a Strouhal number of 47, which can be attributed to the NR presented in Figure~\ref{fig:3DMap_overview}.\\

The background noise spectrum shows a slope similar to the source spectrum. A multitude of tonal components are visible matching often the beamforming spectrum. At Strouhal numbers below 15 two additional peaks can be seen, which were not reconstructed by CLEAN-SC, or are partially on the right wing and thus, are missing in the source spectrum. The additional peak at $\mathit{St}\approx65$ is caused by a model cavity on the right wing. The tone's power is so strong, that it affects the source reconstruction at the given Strouhal numbers on the left wing. At $\mathit{St}\approx30$ the background noise dominates the spectrum and masks the model noise by such an amount that the CLEAN-SC reconstruction fails.\\

\subsection{Aeroacoustic comparison}
In this subsection we analyze the data according to Figure~\ref{fig:datapoints} with respect to the closed and slotted wind tunnel walls in subsection~\ref{sec:SlottedClosed}, with respect to the change in Reynolds number in subsection~\ref{sec:ReynoldsNumberComp}, and with respect to the change in Mach number in subsection~\ref{sec:ReynoldsNumberComp}.

\subsubsection{Closed and slotted test section}\label{sec:SlottedClosed}
\begin{figure}
\centering
\includegraphics[width=\textwidth]{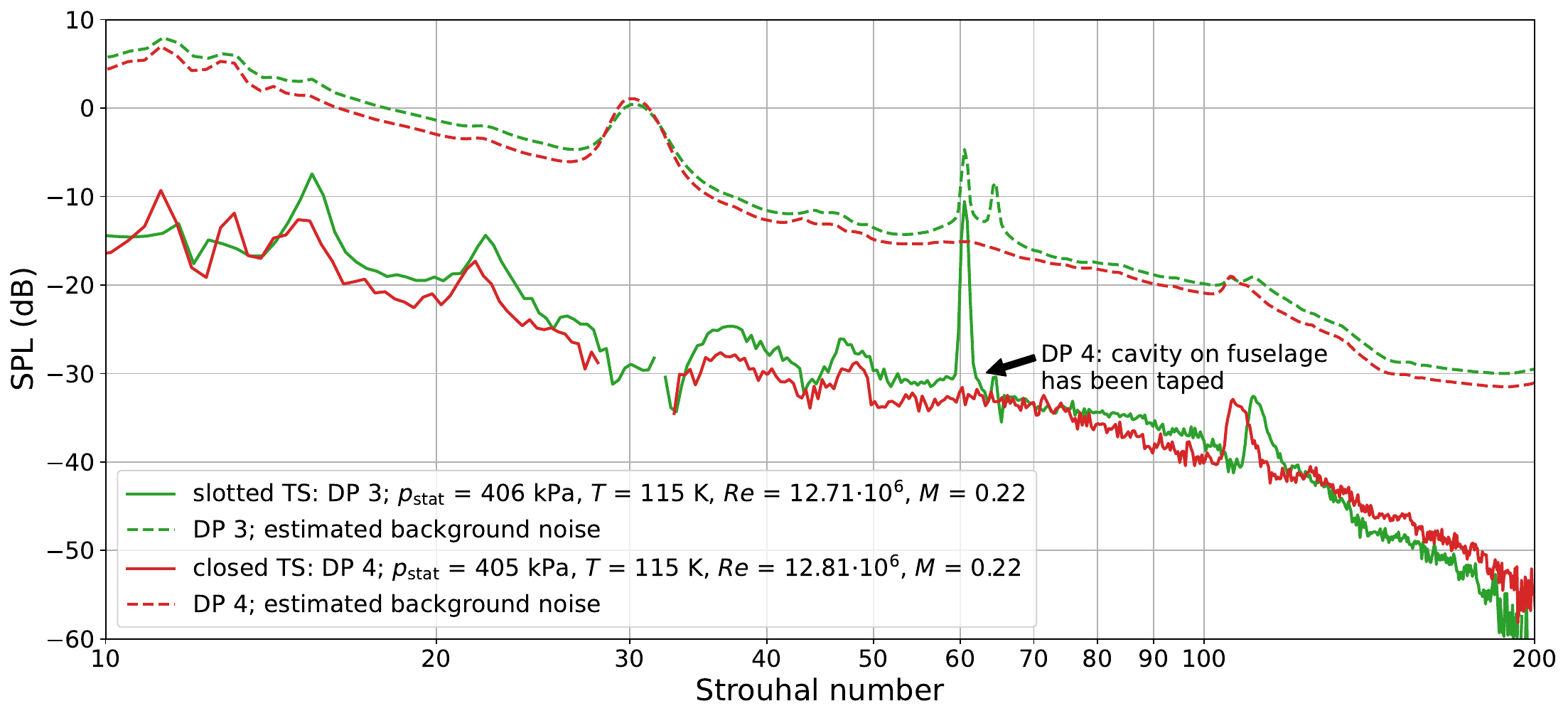}
\caption{Integrated spectra of the left wing, displayed over Strouhal number for DP~3 (slotted TS) and DP~4 (closed TS).}
\label{fig:Spec_vgl_DP3vsDP4}
\end{figure}

Figure~\ref{fig:Spec_vgl_DP3vsDP4} shows the integrated spectra of DP~3 and DP~4, measured at the same global free-stream conditions with slotted and closed test section (TS) walls, see Table~\ref{tab:datapoints}. Additionally, the background noise is shown for both data points, estimated by the mean auto power spectra of all array microphones (using the main diagonal of the CSM). The dominant tone at $\mathit{St}\approx60$ is a model cavity, which was removed at DP~4 and thus, is no effect of the test section walls.\\

For $10<St<60$ the model sound emission within the slotted TS exceeds the closed test section by $\Delta\text{SPL}\approx\SI{2.5}{\decibel}$. Also, the peak Strouhal numbers are increased. Both test sections are subject to the dominant wind tunnel interference at $\mathit{St}\approx30$, see section~\ref{sec:results_sourcespectrum}. At $\mathit{St}\approx70$ the SPL difference becomes neglectable, and then increases towards higher frequencies. The SPL of the NR source at $\mathit{St}\approx110$ is approximately constant while the peak frequency is significantly increased for the slotted TS. The mean auto power of the slotted TS exceeds the one of the closed TS for all frequencies by about $\Delta \text{SPL}\approx\SI{1}{\decibel}$, except for tonal components at $\mathit{St}\approx65$, and the wind tunnel noise at $\mathit{St}\approx30$. In the following comparisons we will only show results of the slotted TS.\\

\subsubsection{Reynolds number variation}\label{sec:ReynoldsNumberComp}
DP~1, DP~2, and DP~3 allow a comparison of the sound emission at constant Mach number $M=0.220$, and increasing Reynolds number, see Table~\ref{tab:datapoints}. Figure~\ref{fig:Maps_vgl_DP1vsDP2vsDP3} shows the corresponding 2D CLEAN-SC maps, integrated over the $z$-axis.\\

For the presented Reynolds numbers, aeroacoustic sources are present at the same regions and the maximum SPL shows small variations of up to $\SI{2}{\decibel}$. At low Strouhal numbers the dominant sources are located in the WLR. Here, small level variations of single sources can be found for different Reynolds numbers. At $\mathit{St}\approx50$ dominant sources are located in the WLR, the NR, and the outer WTR. The level of the source in the NR increases significantly by about 5~dB, but remains constant when further increasing the Reynolds number. At medium Strouhal numbers $\mathit{St}\approx80$ dominant sources are at the inner WTR and the outboard WTR. Here, some of the WLR sources decrease in level with increasing Reynolds number. \\

\begin{figure}
\centering
\includegraphics[width=\textwidth]{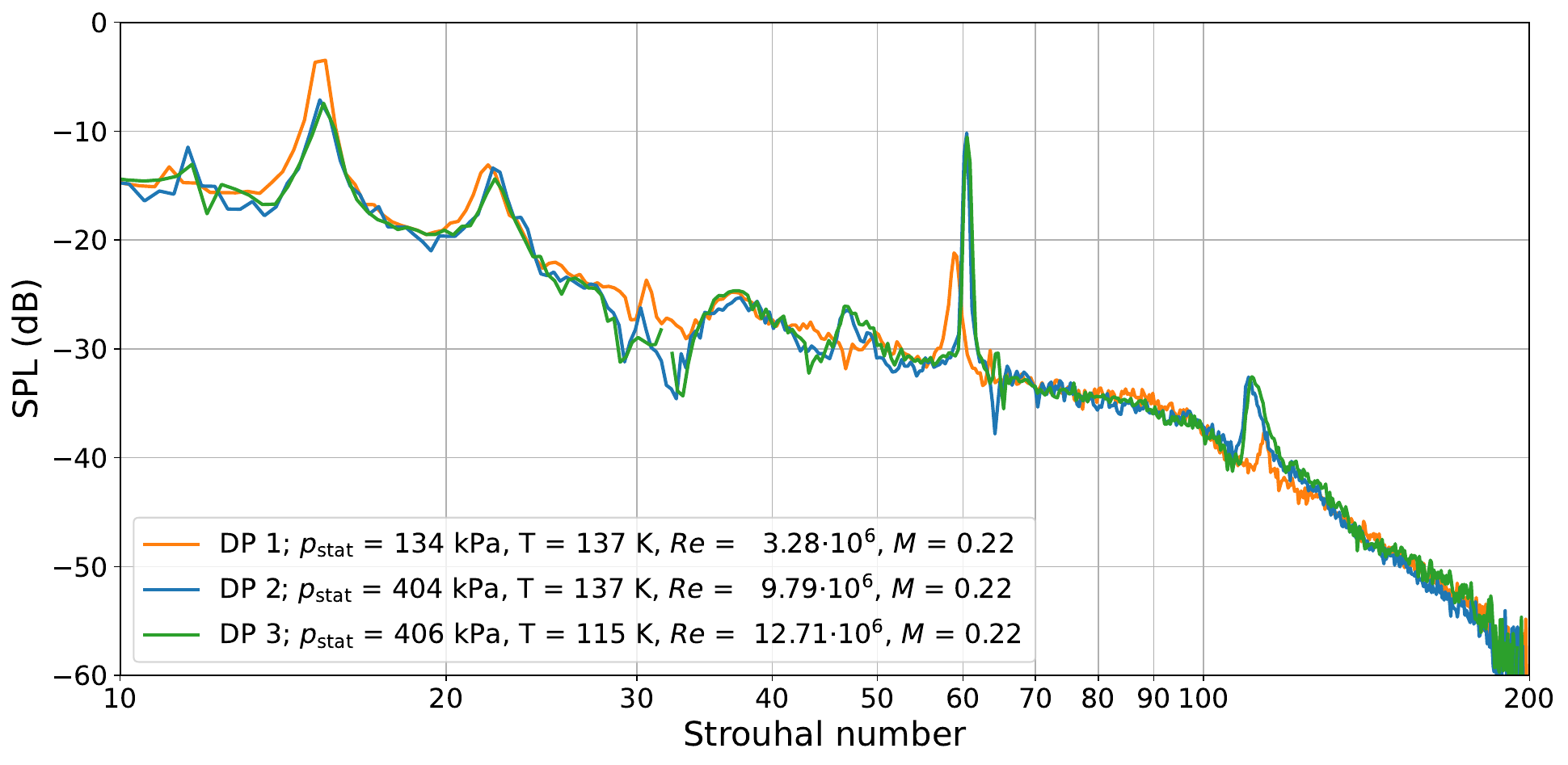}
\caption{Integrated sources of the left wing vs Strouhal number for DP~1, DP~2 and DP~3.}
\label{fig:Spec_vgl_DP1vsDP2vsDP3}
\end{figure}

\begin{figure}
\centering
\includegraphics[width=\textwidth]{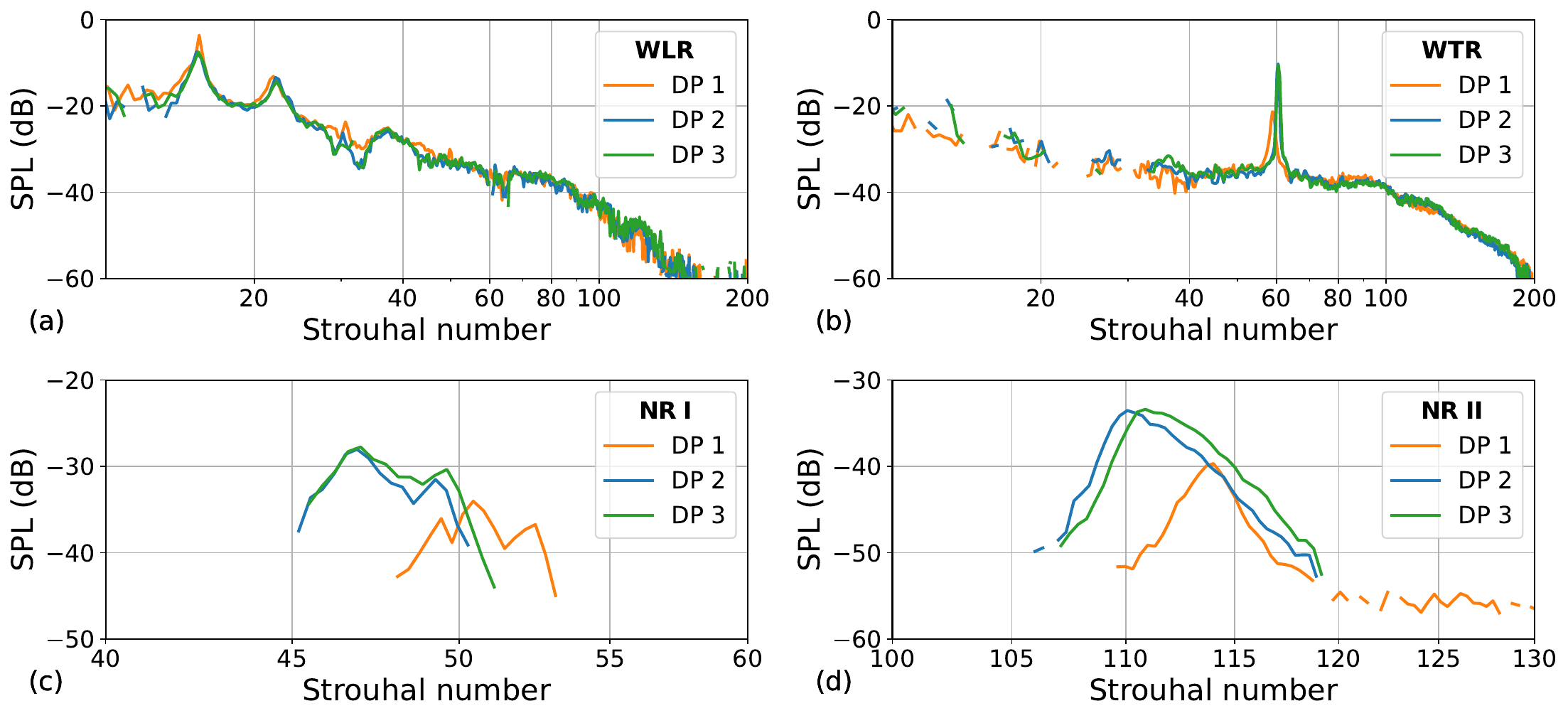}
\caption{Integrated sources of the subgrid volumes (a) WLR, (b) WTR, (c) NR~I and (d) NR~II vs Strouhal number for DP~1, DP~2 and DP~3.}
\label{fig:Spec_vgl_DP1vsDP2vsDP3_sg}
\end{figure}

Figure~\ref{fig:Spec_vgl_DP1vsDP2vsDP3} shows the data points' integrated spectra. Additionally, Figure~\ref{fig:Spec_vgl_DP1vsDP2vsDP3_sg} shows the corresponding ROI spectra, see section~\ref{sec:methods_subgrids}. The broadband SPL of the spectra is very similar, including the microphone calibration correction from section~\ref{sec:miccalib}, and the condition correction from equation~\ref{eq:decibelcorr}. The tonal peaks at $\mathit{St}\approx15$ and $22$ in the wing leading region show an increasing frequency and decreasing SPL with increasing Reynolds number, see Figure~\ref{fig:Spec_vgl_DP1vsDP2vsDP3_sg} (a). At $25<\mathit{St}<35$ the spectrum is insufficiently reconstructed due to the dominating background noise, see section~\ref{sec:results_sourcespectrum}. At $\mathit{St}\approx50$ the source in the NR~I increases in SPL and decreases in frequency with increasing Reynolds number, see Figure~\ref{fig:Spec_vgl_DP1vsDP2vsDP3_sg} (c). At $\mathit{St}\approx60$ the cavity on the fuselage close to the flap increase in SPL and frequency with increasing Reynolds number, see Figure~\ref{fig:Spec_vgl_DP1vsDP2vsDP3_sg} (b). At $\mathit{St}\approx110$ in the NR~II the source increases in SPL and decreases in frequency with increasing Reynolds number, see Figure~\ref{fig:Spec_vgl_DP1vsDP2vsDP3_sg} (d). In the following comparisons we will only show results of the same Reynolds number $\mathit{Re}=10^7$.

\subsubsection{Mach number variation}\label{sec:MachNumberComp}
\begin{figure}
\centering
\includegraphics[width=\textwidth]{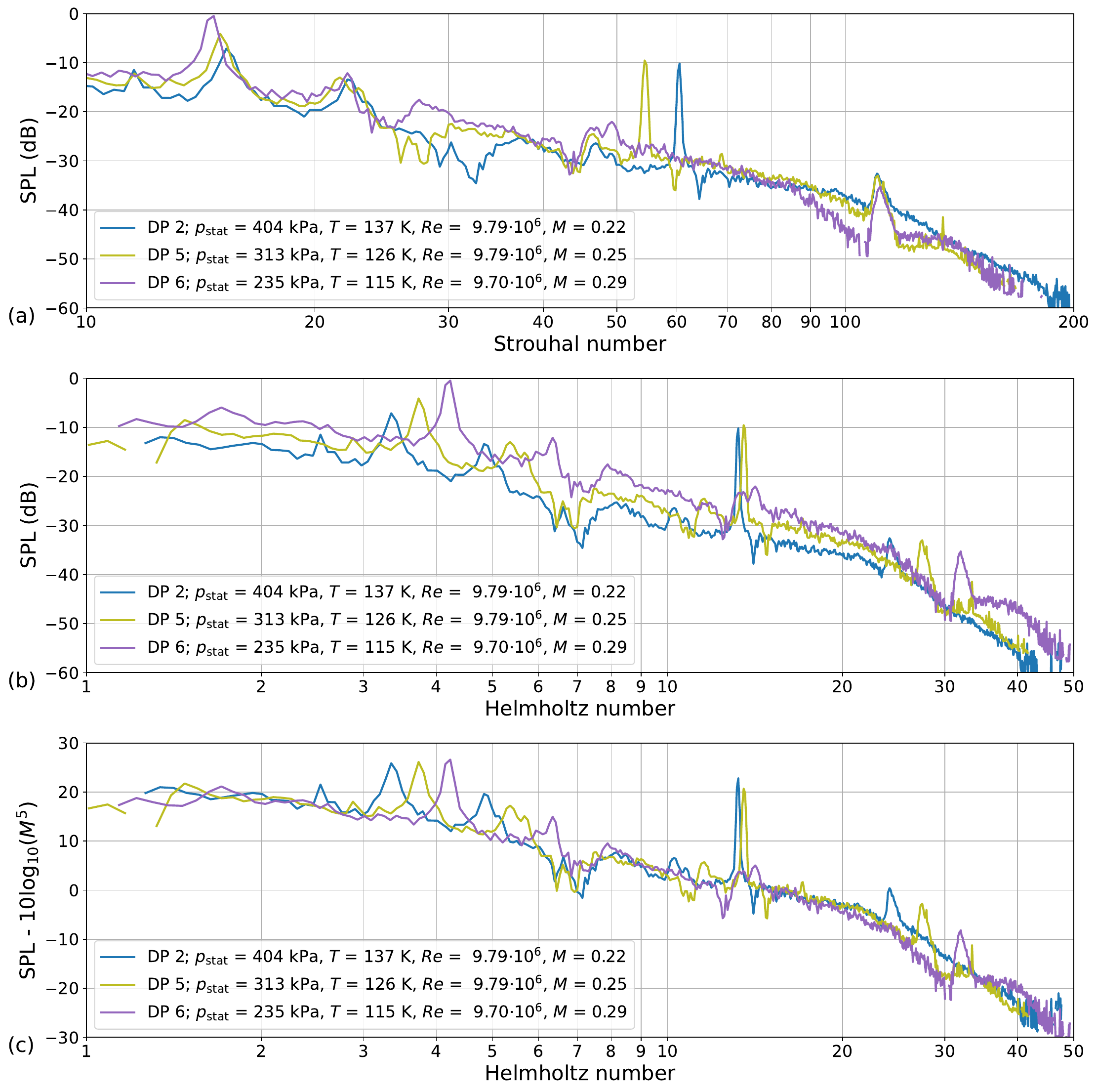}
\caption{Integrated sources for DP~2, DP~5 and DP~6 of the left wing vs (a) Strouhal number and (b) and (c) Helmholtz number. The SPL in the bottom plot is normalized with $M^5$.}
\label{fig:Spec_vgl_DP2vsDP5vsDP6}
\end{figure}

DP~2, DP~5, and DP~6 allow a comparison of the sound emission at increasing Mach number and constant Reynolds number, see Table~\ref{tab:datapoints}. Figure~\ref{fig:Spec_vgl_DP2vsDP5vsDP6} shows the corresponding integrated spectra. Subfigure (a) shows the spectra over the Strouhal number. A reasonable frequency scaling for the WLR tones ($\mathit{St}\approx$~15, and 22) is observed, and a good frequency scaling for the high-frequency NR~II tone ($\mathit{St}\approx$~110). While the WTR tones' power increases with the Mach number, the NR~II tones' power decreases. Subfigure (b) shows the spectra over the Helmholtz number. A reasonable frequency scaling is observed for the cavity tone ($\mathit{He}\approx15$). The cavity tone's power strongly decreases with increasing Mach number. Subfigure (c) shows the scaled SPL, see eq.~\ref{eq:Mach_power_scaling} with $M_0=1$, over the Helmholtz number for a power exponent $n=5$, which is often proposed for airframe noise~\cite{Guo2003}. While the exponent shows reasonable frequency scaling results for the broadband SPL, the various tonal components are not scaled well with this exponent.\\

\begin{figure}
\centering
\includegraphics[width=0.9\textwidth]{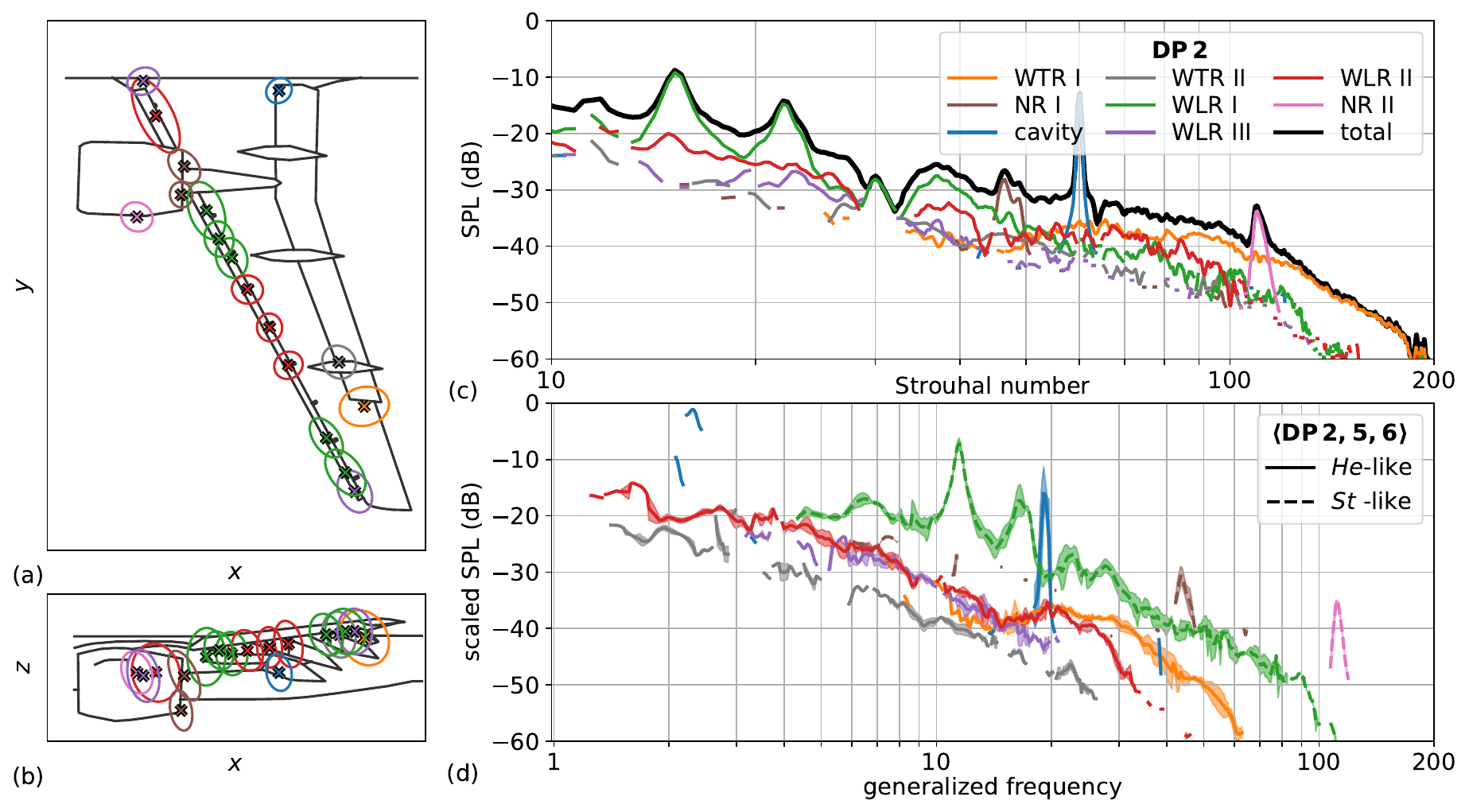}
\caption{The Figure shows source type based 3D-ROI in (a) and (b). Identical colors depict the same source type, for which the integrated spectra of the exemplary DP 08 are shown in (c). For comparison the total integrated SPL from all focus-points on the left wing is depicted in black. (d) shows the averaged spectra with $1\sigma$ standard deviation of DP~2, 5 and 6. The SPL is scaled according to eq.~\ref{eq:Mach_power_scaling} with $M_0=0.220$, and displayed over the generalized frequency, see eq.~\ref{eq:generalized_frequency}. Helmholtz-like scaling spectra are depicted with solid lines, Strouhal-like scaling spectra are depicted with dotted lines. The corresponding scaling-exponents are given in Table~\ref{tab:source_types_parameter}.}
\label{fig:source_types}
\end{figure}

As pointed out in section~\ref{sec:methods_MachVariation} the correct power scaling requires individual source type spectra. Thus, SIND is performed, see section~\ref{sec:methods_subgrids}, to identify individual sources in the 3D beamforming map. The resulting 3D ROI are shown in Figure~\ref{fig:source_types} on a shared $x$-axis and from the (a) $(x,y)$-perspective (bottom), and (b) $(x,z)$-perspective (side). In total 17 individual source ROI are identified. We then group and label similar source types together, based on their spatial occurrence (such as in the WLR) and spectral features (such as tonal peaks), resulting in 8 different source types.~\footnote{As for the prior subsections, the labeling of the sources is based on their spatial occurrences but not on their physical source mechanisms or sound generating object. The naming does not imply a causality between the label and the source mechanism.} The source categories are based on the literature~\cite{Goudarzi2022} and may contain sub- and super groups (e.g., the WLR~II and WLR~III may form a super group). Note, that several source types contain multiple integration ROI, that are not necessarily spatially close to each other such as the WLR~I, and WLR~II. The source type-integrated spectra are shown in Figure~\ref{fig:source_types} (c) for the exemplary DP~2, the corresponding integration ROI in (a) and (b) share the same color.\\

Figure~\ref{fig:source_types} (c) also shows the total SPL (black), integrated from all focus points on the left wing. We observe that the WLR~I sources dominate the SPL at low Strouhal number $10\le \mathit{St} \le 30$, the NR~I and cavity show a prominent peak around $\mathit{St}\approx45$, the NR~II shows a prominent peak around $\mathit{St}\approx110$, and above $\mathit{St}\approx120$ the SPL is dominated by broadband noise from the WTR~I.\\

\begin{table}[h!]
\centering
\begin{tabularx}{\textwidth}{|l||X|X|X|X|X|X|X|X|}

\hline
 & WTR I& WTR II& NR I& NR II& cavity&  WLR I& WLR II & WLR III\\
\hline
$m$& 0.26& 0.03& 0.96& 1.01&  0.25& 0.81& 0.22& 0.28\\
$n$& 2.26& 6.14& 6.40&-1.63&-10.52& 3.89& 4.86& 3.64\\
\hline
\end{tabularx}
\caption{Modification exponent $m$ for the generalized frequency $\widehat{f}$, and Mach power-scaling exponent $n$ for the source types displayed in Figure~\ref{fig:source_types}.}
\label{tab:source_types_parameter}
\end{table}

As pointed out, the sources do not scale perfectly over the free-stream Strouhal or Helmholtz number. This can be attributed to local flow changes at the source positions. Thus, the generalized frequency with optimal modification exponent is employed, see Sec.~\ref{sec:methods_MachVariation} to account for local flow effects. Afterwards, the power scaling exponent is calculated with eq.~\ref{eq:Mach_scaling_exponent} ($\kappa=10$, cf.~\cite{Goudarzi2022}) for each source type. The resulting optimal scaling exponents are given in Table~\ref{tab:source_types_parameter}. Figure~\ref{fig:source_types} (d) shows the resulting Mach-averaged, and SPL-scaled source type spectra, see eq.~\ref{eq:Mach_power_scaling} for $M_0=0.220$, over the generalized frequency, see eq.~\ref{eq:generalized_frequency}. Strouhal-like scaling spectra ($m\ge0.5$) are depicted with dashed lines, and Helmholtz-like scaling spectra ($m<0.5$) are depicted with solid lines. A low standard deviation of the Mach-averaged spectra indicates self-similarity for the given parameters in Table~\ref{tab:source_types_parameter}. Since the standard deviation is low, self-similarity of the sources holds true, which allows an aeroacoustic interpretation of the sources according to Sec.~\ref{sec:SimLaw}.
\\
The self-similar spectra in combination with the dimensionless scaling exponents in Table~\ref{tab:source_types_parameter} provide a quantitative insight into source diagnosis and analysis. For instance, the undesired $He$-scaling tonal artifact with negative Mach scaling exponent suggested a cavity, which was identified and fixed during the measurement campaign, see Figure~\ref{fig:Spec_DP2} and Figure~\ref{fig:Spec_vgl_DP3vsDP4}. In the following section, the identified source types will be acoustically analyzed according to the dimensionless self-similarity laws in Sec.~\ref{sec:SimLaw}. 



%% file: 4_Analysis.tex
The aeroacoustic analysis of the sources are divided into four main categories. First, the spatial occurrence of the sources. Second, the behavior at increasing Mach number and constant Reynolds number. Third, the behavior at increasing Reynolds number and constant Mach number. Forth, the influence of slotted and closed test sections.

\subsection{Spatial occurrence of sources} 
Figure~\ref{fig:source_types} shows, that the main aeroacoustic sources on the left wing are located at the WLR, WTR, and NR. The source locations are mostly symmetric for the right wing, see Figure~\ref{fig:3DMap_overview}, with a slight shift and tilt. This corresponds well with the expected PSF of the array, which is located under and optimized for the left wing. Further investigations are necessary to conclude which differences in location and SPL are explained by the tilted PSF, and which differences can be attributed to source directivity and geometrical masking. Thus, we focus on results from the left wing in this paper.\\

The identified locations coincide with the main sources of comparable models, further referred to as model A~\cite{Ahlefeldt2017} and model B~\cite{Ahlefeldt2013} as shown in the literature~\cite{Goudarzi2022}. Interestingly, for all three models the inner wing trailing regions are silent, or cannot be reconstructed by CLEAN-SC due to the limited dynamic range. A major difference is the WTR~I source, where there are two distinct (leading and trailing edge) sources on model A and model B~\cite{Goudarzi2021,Goudarzi2022,Bai2022} but only one (presumably leading, based on the aeroacoustic behavior) source on the EMBRAER. The positions of the sources remain approximately the same for varying Reynolds number and Mach number, and the slotted and closed test section.\\

\subsection{Mach number dependency}
The present data suggests, that even at constant Reynolds number sources can exhibit a Mach number dependency, based on their aeroacoustic mechanism. Figure~\ref{fig:source_types} and Table~\ref{tab:source_types_parameter} suggest, that there are around eight different source types with different scaling behaviors present on the left wing. The literature~\cite{Goudarzi2022} provides reference values for the generalized frequency modification exponent for several source types of model A and model B.\\

Exemplary reference values are $m=0.79$ for the model D NR~II, and $m=0.95$ for model A NR~II, $m=0.78$ for model B WLR~I, and $m=0.72$ for model A WLR~I. However, these are obtained at increasing Reynolds numbers over increasing Mach numbers, and showed a dependency on the angle of attack. At this point it remains unclear if the modification exponent is constant for source mechanisms within the employed Reynolds number range, as we observed different Mach numbers only for $\mathit{Re}$~$\approx10^7$. For this particular Reynolds number the deviation of the EMBRAER NR~II modification exponent ($m=1.01$) from model A ($m=0.95$) and model B ($m=0.79$) suggests that it is caused by the increase in Reynolds number, and not by the increase in Mach number. The deviation of the EMBRAER WLR~I modification exponent ($m=0.81$) from model A ($m=0.72$) and model B ($m=0.78$) suggests that it is mostly caused by the increase in Mach number. For model A and B the authors identified these as slat cove tones~\cite{Goudarzi2022}. This matches the suggestion that they are caused by Rossiter modes~\cite{Himeno2021}, where the peak-frequencies depend on the convective Mach number at the cove, which might have a non-linear relationship with the free-flow Mach number. Note, that the first Rossiter mode is probably at $\mathit{St}\approx12$ according to the CSM auto spectrum in Figure~\ref{fig:Spec_vgl_DP3vsDP4}, and that the first reconstructed mode at $\mathit{St}\approx15$ is probably the third Rossiter mode. For cavity-noise, a slight increase in the Helmholtz number over Mach number is described in the literature~\cite{Gloerfel2009} for Rossiter modes, matching the experimental observations. For the WLR~II, WLR~III, and WTR~I a similar increase in frequency with increasing Mach number is observed. Note, that the WTR~I source on this wing corresponds to what the authors refer to as a leading flap side edge on model A and model B based on the Helmholtz-like scaling behavior. Trailing flap side edge noise is not present on the EMBRAER wing. Only for the WTR~II a true Helmholtz number scaling is observed. However, due to the small ($\SI{31}{\percent}$), and few variations in Mach number, low modification exponents are subject to some uncertainty. Thus, the WTR~I, cavity, WLR~II, and WLR~III should also be regarded as Helmholtz-like scaling. For the NR~I and NR~II a true Strouhal scaling is observed, which indicates an acoustic mechanism depending on the free-stream Mach number.\\

For the power scaling of aeroacoustic noise the literature proposes $M^5$ for low frequencies~\cite{Guo2003} based on 2D source-mechanisms~\cite{Guo2012}, and $M^6$ for high frequencies because of the dipole character of sources~\cite{Guo2003}. As presented in Figure~\ref{fig:Spec_vgl_DP2vsDP5vsDP6} (c), this gives a reasonable scaling for the broadband SPL. However, the sources that dominate the total broadband SPL, and roughly follow an $M^5$-law are the slat track sources, see Table~\ref{tab:source_types_parameter}. Ffowcs Williams~\cite{FfowcsWilliams1970} proposes that these sources are caused by sharp geometrical edges and also follow an $M^5$-law. Further investigations are necessary to address the issue, which theory is better supported by the data. This should be done using reconstruction algorithms, that preserve spatially extended sources. The slat cove tones are known to exhibit a strong dependence on the angle of attack~\cite{Terracol2016} and dominate the total SPL at lower frequencies and low angles of attack. However, their power scaling is lower than the one of the WLR~II broadband sources. This means that, if present at higher Mach numbers, they will be masked by other sources such as the WLR~II.\\

For the WTR~I and NR~I the scaling exponents exceeds the $M^6$ law, which might be an indication for the dipole source mechanism mentioned before. However, at the observed Mach numbers their SPL contributes little to the total SPL. For sources such as the cavity and the NR~II we observe negative scaling exponents, which means that the SPL decreases with increasing Mach number. The reason for this phenomenon at the cavity might be a growing misalignment of the convective Mach number and the cavity resonance frequency. For the NR~II, the absolute frequency of the tonal peak increases with increasing Mach number, so that the dominant wavelength shrinks compared to the shear layer, which was shown to decrease the SPL and leading to spectral broadening~\cite{Kroeber2013}.\\

Overall, we see that (with exception of the cavity) the standard deviation in Figure~\ref{fig:source_types} (d) is low. Thus, as a function of the source-type dependent generalized frequency, the sources are self-similar for the displayed Mach number range.

\subsection{Reynolds number dependency}
Figure~\ref{fig:Spec_vgl_DP1vsDP2vsDP3} and Figure~\ref{fig:Spec_vgl_DP1vsDP2vsDP3_sg} suggest that within the presented Reynolds number range the aeroacoustic sources maintain the same source mechanism. They transition continuously to higher or lower peak frequencies, and SPL, but the overall shape of the source spectra remains the same. The literature often proposes that multiple acoustic phenomena are observed in wind tunnel experiments which are not present in real-world experiments, due to the mismatched Reynolds number~\cite{Ahlefeldt2013,Ahlefeldt2017} or the relative model geometries and manufacturing tolerances~\cite{Dobrzynski2001}. The presented Reynolds number is close to the full-scale reference Reynolds number. This should ensure that the observed phenomena are comparable to the real-world sound emission. While the differences between low and medium Reynolds numbers (compared to the full-scale reference Reynolds numbers) have a big impact on the aeroacoustic results~\cite{Ahlefeldt2013,Ahlefeldt2017} and lead to different source types on the wing~\cite{Goudarzi2022}, for the EMBRAER model the differences between medium and high Reynolds numbers are small, see Figures~\ref{fig:Spec_vgl_DP1vsDP2vsDP3},~\ref{fig:Spec_vgl_DP1vsDP2vsDP3_sg}. The source types remain the same, however, the peak frequencies and SPL may be under- or overestimated.\\

The Reynolds number dependency of the presented sources is non-linear which makes an estimation or interpolation to real-world Reynolds numbers difficult. E.g., for the WLR~I we mainly observed a strong decrease in SPL of the first reconstructed Rossiter mode ($\mathit{St}\approx15$) when switching from DP~1 to DP~2 but only minor differences when switching from DP~2 to DP~3. Moreover, for the second reconstructed mode ($\mathit{St}\approx22$) and the broadband SPL there are only small variations between the data points.\\

While the cavity on the model (see the tonal component in Figure~\ref{fig:Spec_vgl_DP1vsDP2vsDP3_sg} (b)) was an undesired artifact that was fixed during the measurement campaign, it provides several insights into the nature of cavity noise. For low Re-numbers the thickness of the incoming boundary layer governs the mode selection~\cite{Tam1978}. However, experiments show~\cite{Gloerfel2009,illy2005controle} that at sufficiently high Re-numbers ($\mathit{Re}_L > 10^6$, where $L$ is the depth of the cavity), the thickness of the boundary layer ahead of the cavity has almost no influence on the selection of the oscillation modes. This phenomenon is also reflected by the data (cf. Figure~\ref{fig:Spec_vgl_DP1vsDP2vsDP3_sg}), since for the two higher Reynolds numbers (DP2 and DP3) the SPL and the Strouhal number of the peak caused by the cavity do not change.\\

The NR~II shows a significant dependency on the Reynolds number, which agrees with prior experiments~\cite{Ahlefeldt2013}. However, the presented data do not show the same continuous decrease in Strouhal number with increasing Reynolds number. These observations may be caused by the decreasing boundary layer thickness which exposes larger parts of the strake to the flow $u_{\infty}$. 

\subsection{Slotted and closed test section}
An important task of this measurement campaign was to assess the effects of slotted and closed walls within the test section. While slotted walls offer advantages regarding the wind tunnel's aerodynamic performance~\cite{Meyer2004}, the aeroacoustic disturbance induced by the flow through the slots is rather unknown. Figure~\ref{fig:Spec_vgl_DP3vsDP4} shows that the auto power of the microphones is increased, but so is the reconstructed sound emission below $\mathit{St}<\SI{60}{}$. We therefore assume that the elevated auto power is caused by the increased model sound emission. The reason for this might be a local change in the flow speed and model lift on the model at constant free-flow Mach number and total lift, which is a well-studied phenomenon~\cite{Kroeber2011}. Additional noise from the slots probably appears incoherent due to their varying distance to the microphones, and thus should not affect the source reconstruction. At very low Strouhal numbers $\mathit{St}<\SI{15}{}$ and high Strouhal numbers $\mathit{St}>\SI{110}{}$ the closed test section might offer a slight advantage in the SPL reconstruction. However, the increased SPL might also be related to the local changes of the flow. 

%% file: 5_Conclusion.tex
In this paper we presented the aeroacoustic study of an EMBRAER full model in the cryogenic European Transonic Windtunnel. We presented an end-to-end analysis, from the choice of microphones and measurement parameters, the design of the array and the selection of beamforming and flow parameters, to the source localization, classification and analysis. In particular, the chosen flow parameters allowed to separate the influence of the Reynolds number from the influence of the Mach number on the sound emission. Also, the aeroacoustic influence of slots in the test section was investigated at fixed Reynolds and Mach number.\\

The effect of a slotted test section was small and increased the microphones' average auto powers by around 1~dB. However, the integrated source power was increased by around 2.5~dB at low Strouhal numbers which suggests that the slots do not have a negative effect on the beamforming results at these Strouhal numbers. At very low and high Strouhal numbers a slight advantage of the closed TS was observed. Thus, from an aeroacoustic's viewpoint the slots can be left open for array measurements.\\

The effect of the Reynolds number variation at fixed Mach number showed that there is a strong difference between very low Reynolds numbers, typically achieved in non-cryogenic wind tunnels, and medium Reynolds number (compared to real-world Reynolds numbers). However, the difference between medium and high Reynolds numbers is not that striking, which matches earlier observations by Ahlefeldt~\cite{Ahlefeldt2013}. There are still minor effects, such as spectral broadening and small shifts in peak frequencies and SPL, but we expect the emission results at a full-scale reference Reynolds number to be very similar for the full scale EMBRAER plane.\\

To the authors knowledge, this is the first time that a full model was observed at nearly real-world Reynolds numbers. While the small model scale only achieved $\mathit{Re}_\text{model}/\mathit{Re}_\text{real}\approx0.5$, the large array in comparison provided an excellent resolution. In fact, the depth resolution of the 2D array was sufficient to employ a 3D focus grid. In combination with the SIND method we were able to identify the individual source positions well from the CLEAN-SC results and reconstruct the source spectra even at relatively low full scale frequencies. For future work this opens up new possibilities for further investigation such as the comparison of the left and right wing, and the investigation of source directivities.\\

We showed that the Mach number variation at fixed Reynolds number can have profound impacts on the sources. For our study we used source type dependent ROI and integrated spectra which are necessary to obtain correct scaling parameters. These parameters indicated several different acoustic mechanisms on the wing that are non-linearly related to Mach number. Thus, for a detailed aeroacoustic study we recommend to always separate the Reynolds and Mach number influence. Note, that the parameter estimation strongly benefits from a large amount of measured Mach numbers and a large Mach number range, we therefore recommend to maximize these.\\

For future work, we recommend to validate high Reynolds number wind tunnel measurements with over-flight data, examine source directivities, and use data-fusion with simulated or experimental flow data for the Green's function to further improve and validate the wind tunnel experiments.